\documentclass[conference]{IEEEtran}
\IEEEoverridecommandlockouts
\usepackage{cite}
\usepackage{amsmath,amssymb,amsfonts}
\usepackage{graphicx}
\usepackage{textcomp}
\usepackage{xcolor}
\usepackage{multicol}
\usepackage{balance}
\usepackage{graphicx}
\usepackage{algorithm}
\usepackage{amsthm}

\usepackage[english]{babel}
\usepackage{amsthm}
\usepackage{url}
\usepackage{color}
\usepackage{algorithmicx}
\usepackage{subfigure}
\usepackage{diagbox}
\usepackage{multirow} % Required for multirows
\usepackage[noend]{algpseudocode}
\usepackage{hyperref}
\def\BibTeX{{\rm B\kern-.05em{\sc i\kern-.025em b}\kern-.08em
    T\kern-.1667em\lower.7ex\hbox{E}\kern-.125emX}}
\begin{document}

\title{Brame: Hierarchical Data Management Framework for Cloud-Edge-Device Collaboration\\
}

\author{

\IEEEauthorblockN{1\textsuperscript{st} Xianglong Liu}
\IEEEauthorblockA{\textit{School of Computer Science} \\
\textit{Harbin Institude of Technology}\\
Harbin, China \\
23S003029@stu.hit.edu.cn}

\\

\IEEEauthorblockN{4\textsuperscript{th} Minchong Li}
\IEEEauthorblockA{\textit{School of EECS} \\
\textit{KTH Royal Institute of Technology}\\
Stockholm, Sweden \\
mincoolee@gmail.com}

\and

\IEEEauthorblockN{2\textsuperscript{nd} Hongzhi Wang}
\IEEEauthorblockA{\textit{School of Computer Science} \\
\textit{Harbin Institude of Technology}\\
Harbin, China \\
wangzh@hit.edu.cn}

\\

\IEEEauthorblockN{5\textsuperscript{th} Shenghe Zheng}
\IEEEauthorblockA{\textit{School of Computer Science} \\
\textit{Harbin Institude of Technology}\\
Harbin, China \\
shenghez.zheng@gmail.com}

\and

\IEEEauthorblockN{3\textsuperscript{rd} Yingze Li}
\IEEEauthorblockA{\textit{School of Computer Science} \\
\textit{Harbin Institude of Technology}\\
Harbin, China \\
23B903046@stu.hit.edu.cn}

\\

\IEEEauthorblockN{6\textsuperscript{th} Weihua Sun}
\IEEEauthorblockA{\textit{School of Computer Science} \\
\textit{Harbin Institude of Technology}\\
Harbin, China \\
23S103190@stu.hit.edu.cn}
}
\maketitle

\begin{abstract}
In the realm of big data, cloud-edge-device collaboration is prevalent in industrial scenarios. However, a systematic exploration of the theory and methodologies related to data management in this field is lacking. This paper delves into the sub-problem of data storage and scheduling within cloud-edge-device collaborative environments. Following extensive research and analysis of the characteristics and requirements of data management in cloud-edge collaboration, it is evident that existing studies on hierarchical data management primarily focus on the migration of hot and cold data. Additionally, these studies encounter challenges such as elevated operational and maintenance costs, difficulties in locating data within tiered storage, and intricate metadata management attributable to excessively fine-grained management granularity. These challenges impede the fulfillment of the storage needs in cloud-edge-device collaboration.

To overcome these challenges, we propose a \underline{B}lock-based hie\underline{R}archical d\underline{A}ta \underline{M}anagement fram\underline{E}work, \textbf{Brame}, which advocates for a workload-aware three-tier storage architecture and suggests a shift from using tuples to employing $Blocks$ as the fundamental unit for data management. \textbf{Brame} owns an offline block generation method designed to facilitate efficient block generation and expeditious query routing. Extensive experiments substantiate the superior performance of \textbf{Brame}. 
\end{abstract}


\section{Introduction} \label{sec:intro}
Cloud-edge-device collaboration (\textbf{\underline{CEDC}}) architecture widely exist in various areas such as intelligent manufacturing and smart grids~\cite{Liao2022CloudEdgeEndCI}~\cite{Zhang2022BlockchainAF}. Data management plays a crucial role in \textbf{CEDC}~\cite{Fan2023MSIAPAD}~\cite{Li2022ApplicationDrivenDM}. Unfortunately, this area is seldom studied in the community~\cite{Cui2023TSCabinetHS}, and the straightforward application of existing data management techniques lose the optimization chance for \textbf{CEDC}~\cite{Li2018AnEM}~\cite{Tao2020SecureDS}. Thus, effective data management techniques for \textbf{CEDC} are in great demand.

The major challenge of data management for \textbf{CEDC} is that data may be located in one or more sides of cloud, edge and end, and need to be arranged subtly to achieve high performance without losing consistency. 
 
Cloud servers possess powerful computing and storage capabilities but elevated query latency, making them ideal for historical data storage. In contrast, Edge databases have limited computing and storage but offer lower query latency, mainly serving as primary caches for cloud servers to store local hot data. Terminal devices, while constrained in computing and storage capabilities, capitalize on their proximity to end-users to fulfill real-time query requirements. The disparate computing and storage capabilities of \textbf{CEDC} devices necessitate assigning distinct storage roles to them. Establishing a multi-level cache between cloud servers and end-users optimizes data access, enabling efficient collaboration.

Thus, it is natural to design a hierarchical storage structure for \textbf{CEDC}, However, prevailing hierarchical data storage and scheduling architectures are not explicitly crafted for \textbf{CEDC}. Their primary emphasis lies in formulating effective mechanisms for discerning and isolating hot and cold data, subsequently organizing them into distinct layers, commonly referred to as temperature-based hierarchical storage. While this paradigm has found widespread application in scenarios such as cloud-edge and cloud-device data tiering storage~\cite{Li2018AnEM}~\cite{Lin2019ATD}, its adaptation and implementation in the realm of \textbf{CEDC} pose critical challenges that demand immediate attention.

\textbf{C1: How to vertically expand the hierarchy of data storage within the table?} 
In \textbf{CEDC}, different tiers exhibit distinct storage requirements. Drawing on the prevalent three-tier architecture of DaaS providers (Cloud) - enterprise local servers (Edge) - enterprise employees (End) as an illustrative example. Edge servers must fulfill the needs of enterprise employees, necessitating stored data to reflect the collective needs of multiple users over time. Conversely, user devices situated at the end need only deliver low processing latency services for personalized queries originating from local users. When extending the temperature-based two-tier storage strategy~\cite{Levandoski2013IdentifyingHA}~\cite{Pelkonen2015GorillaAF} to the three-tier architecture of \textbf{CEDC}, it is crucial not to overlook the differentiated storage requirements among different tiers.

\textbf{C2: How to reduce the overhead of maintaining fine-grained statistical information within the cross-tier table?} 

The statistical approaches~\cite{Levandoski2013IdentifyingHA}~\cite{Pathak2018LifeCO}~\cite{Cui2023TSCabinetHS} to gauge the hotness and coldness of data result in substantial computational and maintenance costs. These methods necessitate the ongoing upkeep of temperature and other statistical information for each tuple, with frequent updates of the maintained numerical information. Additionally, the time cost of retrieving low-temperature data and sorting high-temperature data cannot be overlooked during data migration.

\textbf{C3: How to preserve data locality within a tier?}
Methods distinguishing between hot and cold data using data structures~\cite{Hsieh2006EfficientIO}~\cite{Park2011HotDI} may disrupt the positional relationships between data tuples. Such methods rely on the location characteristics of the tuple in the data structures to represent their hot/cold status. When data is distributed across different storage devices based on temperature, the constraints on the integrity of query results may lead to additional data location costs. Since without sophisticated data locality mechinism, tuples satisfying the query conditions may be located on any device, and for precise queries, it has to scan more data to avoid missing results.

\textbf{C4: How to reduce the cost of metadata management?} 
Upon receiving a query, the data management system depends on metadata, mainly Max-Min indexes, to ascertain the required data pages for the query. When tuples are exclusively organized into pages based on temperature, there exists a probability of metadata-based index failure, leading to scanning unnecessary pages. Moreover, when data is dispersed across multiple storage devices, queries may be erroneously routed to higher-level storage devices. 
Thus, when data is migrated at the tuple level, it is necessary to reorganize and generate metadata information for tuples placed on the same-tier devices. For scenarios involving frequent migration, such re-construction overhead is high.

We find two basic problems leading to the challenges:

\textbf{Excessive granularity in data management}: The measurement of data temperature at the tuple level inevitably introduces significant computational overhead and poses difficulties in data localization. Specifically, the approach of maintaining temperature attribute for each tuple and updating them using a workload-aware method becomes impractical in scenarios involving massive data. In cases where the storage location of tuples changes, it becomes necessary to reorganize data belonging to the same tier and generate metadata for localization. This incurs an unpredictable metadata management cost, particularly in scenarios with frequent data migration.

\textbf{Inapplicability of classical temperature-based data migration architectures to \textbf{CEDC}}: On the one hand, current research predominantly focuses on designing mechanisms for identifying and segregating hot and cold data, storing them in separate layers. However, this approach restricts the vertical scalability of storage hierarchies. On the other hand, there is a noticeable gap in research aiming at establishing a unified framework to cater to the diverse requirements of hierarchical data storage in the context of \textbf{CEDC}.

\begin{figure}
  \centering
  \includegraphics[width=\linewidth]{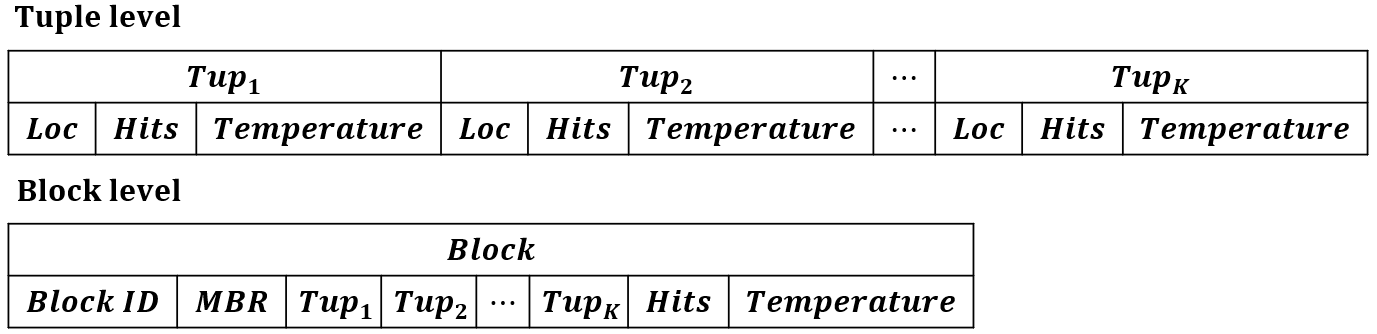}
  \caption{Tuple-level vs Block-level.}
  \label{fig:struc}
  \vspace{-2.0em}
\end{figure}

To enable effective \textbf{CEDC}, we attempt to solve the above two issues by slightly increasing data management granularity and introducing a three-tier data management architecture. For the former, we suggest using data blocks ($Blocks$) as the basic units for data management, where each $Block$ is a set of data tuples with a predetermined maximum capacity. \autoref{fig:struc} illustrates the structural differences between using tuples and $Blocks$ as basic units. We generate $Blocks$ with a workload-aware unsupervised method (see Section~\ref{sec:bg} for more details). For the latter, we allocate distinct storage roles to cloud, edge, and terminal devices, catering to diverse requirements in \textbf{CEDC} storage, with $Block$ as the smallest granularity (Details can be found in Section~\ref{sec:dm}).
Integrating the aforementioned designs, we propose \textbf{Brame}, a \underline{B}lock-based hie\underline{R}archical d\underline{A}ta \underline{M}anagement fram\underline{E}work for \textbf{CEDC}.

In conclusion, this work makes the following contributions:

1. We propose \textbf{Brame}, a three-tier data storage architecture customized for \textbf{CEDC}, assigning distinct storage roles and functions to cloud, edge, and terminal-user devices. This effectively addresses a gap in the current research landscape.

2. To address limitations in existing research, \textbf{Brame} takes $Blocks$ as the basic unit for data management. \textbf{Brame} is equipped with a workload-aware data reorganization technique to consolidate tuples with similar query access patterns into the same $Blocks$. Moreover, we develop a workload-driven table partition algorithm for \textbf{Brame} to make this data reorganization technique suitable for scenarios with massive data.

3. We conduct experiments using two real datasets on two downstream tasks to evaluate \textbf{Brame}. The results demonstrate the superior performance of the proposed techniques.

The subsequent structure of this paper unfolds as follows. We propose a problem statement in ~\ref{sec:stm}. Section~\ref{sec:sys} outlines the comprehensive architectural design and workflow of \textbf{Brame}. Section~\ref{sec:bg} delves into the block generation approach, which comprises table partition and intra-table data reorganization components. Section~\ref{sec:dm} introduces the collaborative three-tier data storage scheduling strategy for \textbf{CEDC}, centered around $Blocks$. Section~\ref{sec:exp} coducts a set of thorough experiments aimed at showcasing the superior performance of \textbf{Brame}. In Section~\ref{sec:rel}, we conduct a review of pertinent literature in the field of data storage and management. Lastly, Section~\ref{sec:con} provides a summary of our work.

\section{Preliminary} \label{sec:stm}

\textbf{Scope}: We focus on organizing and placing data within a single table in relational databases, which is crucial for \textbf{CEDC}. In real-world business environments, tables can contain billions of tuples, necessitating partitioning and distribution across multiple databases. Frequently accessed data should be placed on edge or terminal devices to reduce latency and improve performance.

Moreover, since multi-table joins are costly, single-table queries dominate in such scenarios, with multi-table queries being facilitated by multiple single-table queries. As an early work in this field, our research focuses on how to meet the storage needs of \textbf{CEDC} by employing a data organization and tiered storage strategy without replicas, ensuring efficient data access and management.

\textbf{Query}: In this paper, we focus on read-only queries of the following form:
\textit{Select  $*$  From T Where  $l_1 < col_1 < u_1$ and \ldots and  $l_k < col_k < u_k$}

In future research, we plan to gradually introduce support for operators such as Join, Group By, and Order By.

\textbf{Data Place and Migration}: Given a table $T$, tuples within the table are organized into $Blocks$ according to certain rules, with the $Block$ being the smallest unit of storage management. The logical storage format of a $Block$ is illustrated in \autoref{fig:struc}. For a query $q$, we determine whether to access a $Block$ based on its metadata. In a \textbf{CEDC} scenario, $Blocks$ are placed in different locations based on their temperature; the closer a $Block$ is placed to the end side, the lower the query access latency. For data placement tasks, we aim to position frequently accessed $Blocks$ closer to the edge and end sides near the end users, while colder $Blocks$ are placed in the cloud. Since query access patterns from the end side are often time-varying, we aim to schedule $Blocks$ based on the current workload pattern. This scheduling can be real-time or periodically executed:

\textbf{Periodic data scheduling}: Given a workload \( W = \{ q_1, q_2, \ldots, q_m \} \) and a set of $Blocks$ \( B = \{ b_1, b_2, \ldots, b_n \} \), determine the placement strategy for $Blocks$ to maximize the hit rate of queries at the edge and end sides.

\textbf{Real-time cache replacement}: Given a query \( q \) and a set of $Blocks$ $B$, determine which $Blocks$ \( B = \{ b_1, b_2, \ldots, b_n \} \) to evict from the terminal device side to the cloud or edge side based on the $Blocks$ required by \( q \).

\section{Framework} \label{sec:sys}

\begin{figure}
  \centering
  \includegraphics[width=\linewidth]{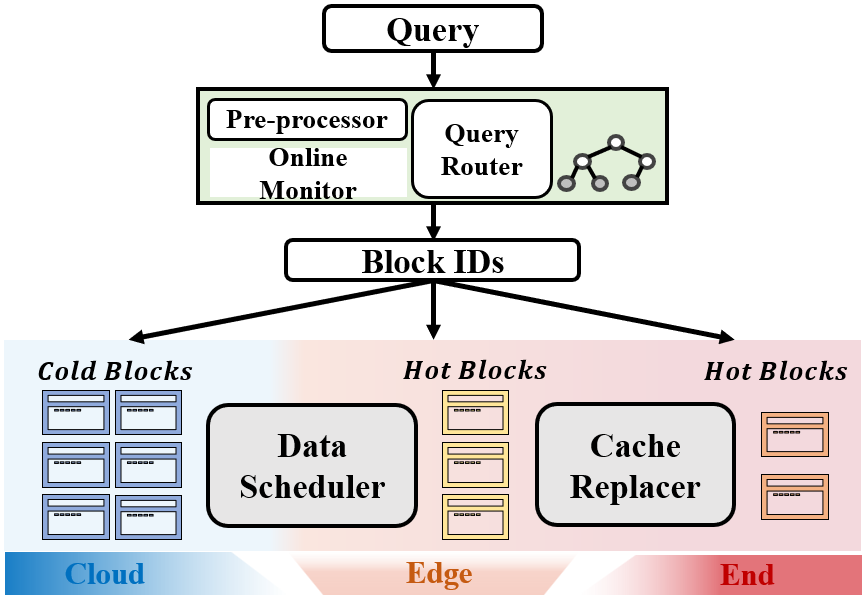}
  \caption{An Overview of \textbf{Brame}.}
  \label{fig:system}
  \vspace{-2.0em}
\end{figure}

\textbf{Brame} is a comprehensive hierarchical data management system designed for \textbf{CEDC}, incorporating a variety of technologies. The overarching design philosophy centers around the adoption of $Blocks$ over tuples as the fundamental unit for data storage and scheduling. This is achieved by grouping tuples that exhibit similar query access patterns within the same $Block$.
Furthermore, to meet the demands of data migration, we develop online monitoring algorithms for real-time queries to identify currently hot $Blocks$ for \textbf{Brame}. Generally speaking, \textbf{Brame} has two distinct and decoupled modules: the offline block generation (\textbf{\underline{BG}}) component and the online hierarchical storage scheduler (\textbf{\underline{HSS}}) component. The former partitions the original table into $Blocks$ in a workload-aware way. The latter performs online block scheduling. \textbf{Brame}'s architecture is depicted in \autoref{fig:system}.

\begin{figure}
  \centering
  \includegraphics[width=\linewidth]{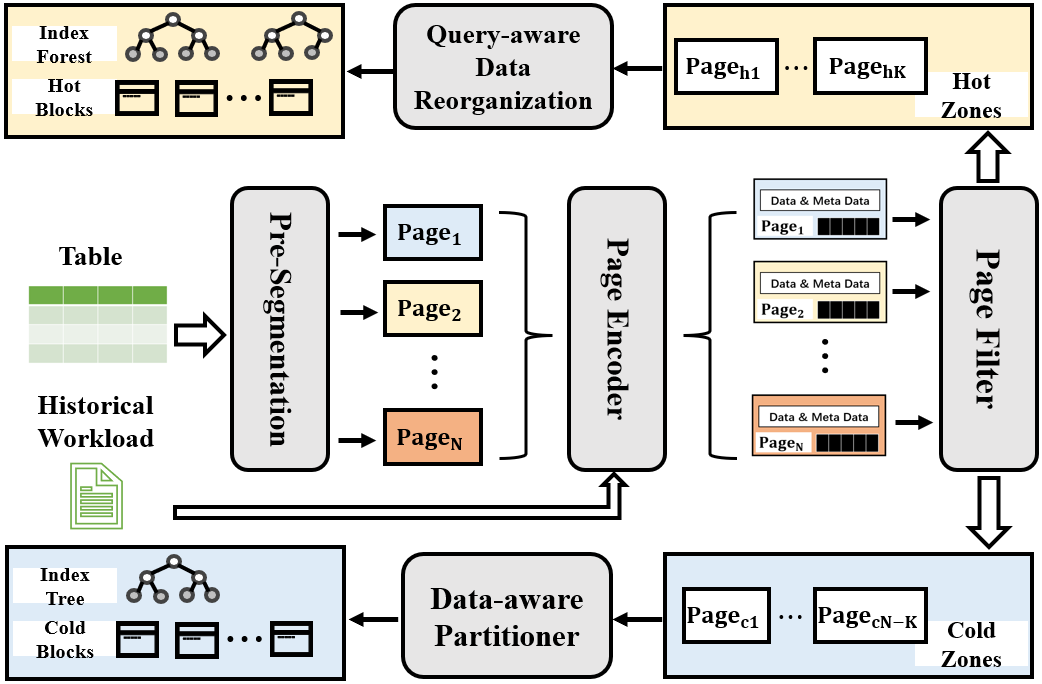}
  \caption{The Workflow of \textbf{Brame}'s Offline Block Generation Technology.}
  \label{fig:bg}
  \vspace{-2.0em}
\end{figure}

\parindent=9pt
The \textbf{BG} module, as illustrated in \autoref{fig:bg}, serving as an external control component for cloud databases, is deployed within the cloud side. Given the table $T$ and a representative workload $W$, \textbf{BG} organizes tuples into $Blocks$ in a workload-aware manner, involving two primary operations: \textit{table partitioning} and \textit{intra-table data reorganization}. Specifically, for $T$ migrated to the cloud, \textbf{BG} instructs the cloud database to partition the table into small data pages and establish feature vectors for these pages based on workload $W$. Infrequently accessed pages are treated as cold data and filtered out by $BG$. For the remaining pages, $BG$ employs the K-Means method with balance constraints~\cite{Liu2018FastCW} to cluster them, with each cluster corresponding to a sub-table derived from $T$. 

Once the table partitioning is complete, \textbf{BG} reorganizes the data within each subtable in parallel. This involves acquiring feature encodings for the data tuples within the sub-table in a workload-aware manner. Subsequently, the HBC method, whose details will be introduced in Section~\ref{subsec:dr}, is used iteratively to partition large clusters into smaller clusters of similar size, until all clusters can comfortably fit into a $Block$. The tuples within each $Block$ are arranged in a user-defined order (e.g., lexicographical order).

Furthermore, to facilitate efficient data location and query routing, \textbf{BG} instructs the cloud database to consolidate tuples from pages that have been filtered out by the data partitioning component. Subsequently, the amalgamated data is structured into $Blocks$ using the K-D Tree framework~\cite{Friedman1976AnAF} and stored in the cloud. The K-D Tree structure, due to its simplicity and effectiveness, as well as query-friendliness, serving as an index for cold data, is duplicated by \textbf{BG} for deployment on edge and end devices.
For the $Blocks$ generated by the data reorganization component, \textbf{BG} disseminates the intermediate outcome of the HBC algorithm process, known as the hierarchical clustering tree, as the query routing structure to edge and end devices. The root node of the hierarchical clustering tree is employed to establish its connection with a specific sub-table, while intermediate nodes retain metadata for the subordinate data tuples, including Max-Min index. Leaf nodes record metadata pertaining to the corresponding $Block$ and its placement location. 
Consequently, when a query is received, it determines the $Blocks$ it requires access to by scanning the K-D Tree for cold data and the hierarchical clustering tree for high-frequency data. Subsequently, it forwards read requests to the appropriate storage devices.

\textbf{HSS}, deployed at the edge and the end, acting as a scheduler for data migration. It determines the placement and migration strategy of Blocks generated by the \textbf{BG} component within the \textbf{CEDC} framework. \textbf{HSS} actively monitors query processing requests initiated from the end. When a query arrives, it enters the pre-processor module of \textbf{HSS}, which identifies the $Blocks$ need to access using the query routing strategy described earlier. In the context of collaborative cloud-edge data-tiering storage, \textbf{HSS} periodically activates the temperature calculation module and data migration module. The temperature calculation module assesses the importance of each $Block$ in the current time period based on the current and historical access frequency features. It then quantitatively calculates the temperature of each $Block$ using the temperature model, where the temperature reflects the potential benefits of storing it at the edge. The data migration module models the problem of collaborative cloud-edge data-tiering placement as a \textit{0-1 knapsack problem} ~\cite{Du2011DesignAA} to determine the hierarchical block placement scheme. It instructs the edge DBMS to migrate data in batches. At the same time, the storage location information of relevant $Blocks$ in the K-D Tree and hierarchical clustering tree(forest) is updated. For terminal devices, which primarily serve personalized query processing needs for local users, \textbf{HSS} delegates the responsibility of data migration scheduling to cache replacement strategies. These strategies determine which $Blocks$ should be cached locally.

We will introduce \textbf{BS} and \textbf{HSS}  in Section~\ref{sec:bg} and Section~\ref{sec:dm}, respectively.

\section{Block Generation} \label{sec:bg}

Since \textbf{Brame} employs $Blocks$ as the fundamental unit for storage scheduling, the quality of block generation significantly influences the system's overall performance. This section delves into the offline block generation issue, with particular emphasis on three key aspects: (1) the benefits of increasing the granularity of data management from tuple level to $Block$ level in Section~\ref{subsec:bgm}, (2) the strategy for organizing data that is alike or exhibits analogous query access patterns within a limited number of $Blocks$ in Section~\ref{subsec:dr}, and (3) the approaches employed to ensure the adaptability of our method in handling extensive datasets in Section~\ref{subsec:tp}.

\vspace{-0.5em}
\subsection{Benefits} \label{subsec:bgm}

Logically, a $Block$ is a group of data tuples with a predetermined maximum capacity for tuples. \autoref{fig:struc} shows the key structural differences between using tuples and $Blocks$ as the basic storage units.
For data management at the tuple level, it becomes imperative to preserve temperature and access frequency of subsequent temperature updates for each tuple over a defined temporal window. Additionally, when data migration occurs, it is necessary to reorganize data belonging to the same layer and generate metadata for localization. Raising the granularity of management to the $Block$ level mitigates the aforementioned issues. We only need to maintain and update temperature and access frequency information for each $Block$, rather than individual tuples within the $Block$. Additionally, when conducting data migration at the $Block$ level, the tuples within a $Block$ are treated as a cohesive unit. This design choice guarantees the preservation of the locality of tuples within a $Block$, obviating the need for metadata reconstruction. During query execution, leveraging the MBR(Max-Min Index) of each $Block$ aids in determining the $Blocks$ to be scanned, subsequently identifying tuples that satisfy the query constraints.

\vspace{-0.5em}
\subsection{Data Reorganization} \label{subsec:dr}

Before formally introducing our data reorganization approach, we use a simple example to demonstrate the motivations of the workload-aware $Blocks$ generation approach.

\begin{figure}
  \centering
  \includegraphics[width=\linewidth]{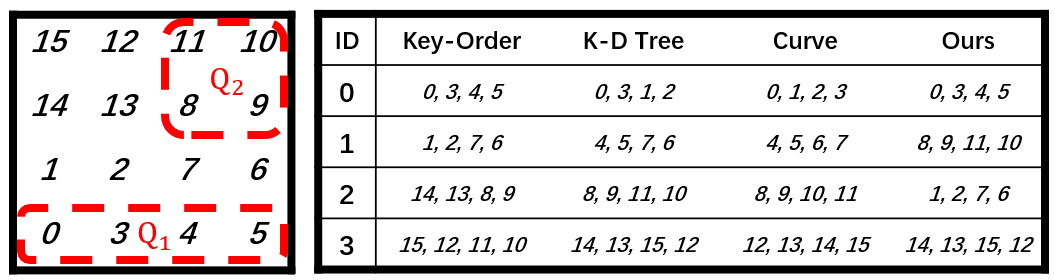}
  \caption{Workload-aware vs Data-aware Approaches.}
  \label{fig:eg1}
  \vspace{-2.0em}
\end{figure}

\textit{\textbf{Example}: Given a 2-dimensional table $T$ with 16 tuples and a workload $W=\{Q_1, Q_2\}$, \autoref{fig:eg1} illustrates the differences between three data-aware methods and the workload-aware method in constructing $Blocks$ with a $Block\_size$ of 4.}

\textit{In our approach, tuples 0,3,4,5 are accessed exclusively by query $Q_1$, indicating the same query access pattern. Consequently, our workload-aware strategy groups them into a single $Block$. Similarly, tuples 8,9,11,10 are consolidated into one $Block$. For the remaining tuples, despite sharing a query access pattern, they are split into two $Blocks$ due to the block size constraint.}

\textit{Taking periodic data migration as an example, if we set the cache budget to 2 ($Blocks$), during a migration execution cycle, when the test workload is the same as $W$, only our method can load all the required tuples into the cache. Similar results are observed in real-time cache replacement.}

In order to efficiently consolidate tuples with similar or analogous workload access patterns into a single $Block$ or a minimal number of $Blocks$, we propose a workload-driven data reorganization algorithm. This algorithm centers its analysis on data tuples as fundamental units and groups them based on their similarity, resulting in a more concise data representation. The central objective of this algorithm is to establish a means of quantifying the similarity between tuples. Our intuitive approach suggests employing the queries served by tuples as features for similarity analysis: when two tuples serve the same workload, they exhibit the highest degree of similarity and are consequently grouped within the same $Block$.
With this insight, we propose the following workload-aware tuple reorganization approach.

\textbf{Workload-aware Tuple Encoding}: Given a table $D$ and a set of representative queries $W={\{q_{1}, q_{2}, \ldots, q_{|W|}}\}$, for each $q \in W$, we scan $D$ to obtain a bitmap representation, donated as $bitmap(q,D)$ for each query $q$. These bitmap representations are then concatenated these representations to construct the feature matrix $X=[bitmap(q_{1},D)||\ldots||bitmap(q_{|W|},D)] \in \{0,1\}^{|D| \times |W|}$. The $i$th row of $X$ represents the feature vector of the $i$th tuple.  

In devising the restructuring plan for data, we leverage the encoding representation of tuples, addressing a typical unsupervised problem. We utilize the K-Means clustering method for tuple reorganization. While the K-Means method proves more accommodating for clustering large-scale data compared to other classical clustering techniques like DBSCAN~\cite{Ester1996ADA} and hierarchical clustering~\cite{Ward1963HierarchicalGT}, and it does not impose specific requirements on the data distribution~\cite{Jain1999DataCA}, our practical experience reveals challenges in directly applying this traditional clustering method to Block generation. The difficulties stem from the following reasons:

1.Sensitivity to the choice of $K$, and a lack of control over cluster size, resulting in clusters that can be excessively large or too small.

2.Difficulty in facilitating efficient query routing. The establishment of effective indexing structures for each $Block$ necessitates additional overhead.

3.High time complexity. traditional K-Means methods exhibit a time complexity of $O(|D| \times K \times E)$, making them unsuitable for clustering extensive datasets.

To address these challenges, we develop a novel approach, Hierarchical Iterative K-Means clustering algorithm (hereafter referred to as HIKM), the basic idea is to iteratively partition large clusters into a few smaller clusters of similar scale in a top-down, hierarchical clustering manner, until each cluster can fit into a $Block$.

\textbf{Hierarchical Iterative K-Means (HIKM)}: 
Our target is to organize tuples with similar encodings together while controlling the size of each cluster. To achieve this goal, HIKM provides a viable optimization approach. This method progressively optimizes the organization of data tuples in a hierarchical manner. During the clustering process, we dynamically adjust the value of $k$, based on the current cluster size and the characteristics of the tuples within the cluster.

When the cluster count $k$ is fixed, the time complexity of the Hierarchical Iterative K-Means clustering algorithm is $O(|D| \times k \times log_k(K) \times E)$, offering an improved computational efficiency compared to traditional K-Means algorithms.

Throughout the execution of our HIKM algorithm, we maintain and update a hierarchical clustering tree to facilitate efficient query routing. In particular, for each sizable cluster that can be divided, we treat it as an intermediate node within the tree. We also establish and maintain Max-Min index based on the numerical features of the tuples within the cluster. Subsequently, we employ the K-Means algorithm to split it into a series of smaller clusters, which are recorded and saved as child nodes of that intermediate node. Non-divisible clusters, or $Blocks$, correspond to the leaf nodes of this tree. Consequently, when a query is received, we can expeditiously identify the requisite $Blocks$ for access based on the structural attributes of the hierarchical clustering tree. The use of Max-Min index for query routing ensures that no data is overlooked.

Additionally, we introduce a balancing constraint into the K-Means algorithm~\cite{Liu2018FastCW} to promote HIKM by subdividing extensive clusters into smaller clusters of similar sizes as it employs the K-Means iteratively within its hierarchical structure.

\textbf{Balanced K-Means}: Given a cluster $\hat{D}$ and its corresponding feature matrix $\hat{X}$, along with the number of clusters $\hat{K}$, the primary objective of balanced clustering is to explore the inherent structure of the data and yield clusters of approximately equal sizes. The optimization objective is as follows:
\begin{equation}
\mathop{\text{min}}\limits_{\hat{H},\hat{C}} ||\hat{X} - \hat{H} \hat{C}||^2_F + \varphi \sum_{i=1}^{\hat{K}} (\sum_{l=1}^{|\hat{D}|}|\hat{H}_{l,i} - |\hat{D}|/\hat{K})\textsuperscript{2},  s.t. \sum_{i=1}^{\hat{K}}\hat{H}_{l,i} = 1
\end{equation}
Here, $\hat{C}$ stands for the cluster centroids, $\hat{H}$ is the cluster assignment matrix, and the hyperparameter $\varphi$ symbolizes a trade-off between clustering and balance constraints. We incorporate the balancing constraint into the K-Means method with the intention of equalizing the cluster sizes. On the one hand, we prevent the formation of small clusters leading to high block vacancy rates. On the other hand, we mitigate the emergence of large clusters that could cause profound bifurcations in the hierarchical clustering tree. Such bifurcations result in degradation and imbalance issues, negatively impacting query search efficiency.

SPANN~\cite{Chen2021SPANNHB} also employs similar techniques to address other issues within the realm of databases. Adopting a consistent nomenclature with SPANN, we name the HIKM algorithm incorporating balance constraints as Hierarchical Balanced Clustering (HBC).

\textbf{Incorporation of Data-aware Partition Technology}: In the practical implementation of the hierarchical clustering algorithm, we observe the skewness of the feature vectors for the tuples to encode in intermediate nodes. In extreme cases, these feature vectors within a node might be identical, this may result in clustering failure or the formation of highly imbalanced cluster partitions. In such scenarios, we utilize a data-aware approach like K-D Tree to partition the tuples within this node, generating $Blocks$.

We combine the utilization of the three aforementioned techniques for data reorganization, as illustrated in Algorithm~\ref{alg:dr}.

Algorithm~\ref{alg:dr} and ~\ref{alg:tp} in Section~\ref{sec:bg} together form the basic algorithm for the BS component, while the fundamental algorithm for HSS is Algorithm~\ref{alg:bm} in Section\ref{sec:dm}. %Due to space constraints, the combination of Algorithm 1 and 2 is not presented.

In Algorithm~\ref{alg:dr}, we encapsulate our data reorganization method. In the first step of the algorithm, we employ workload-aware tuple encoding technology to encode tuples. Subsequently, we iteratively subdivide extensive clusters into smaller clusters of comparable sizes using the HBC algorithm (Lines 4-15). Throughout the execution of the HBC algorithm, we uphold and revise the hierarchical clustering tree (Line 9). Once the data reorganization is concluded, we establish the initial placement of $Blocks$ at the edge-cloud boundary based on representative workloads (Line 16). For a more comprehensive understanding, please refer Section~\ref{sec:dm}.

The time complexity of the HBC clustering algorithm is $O(|D| \times k \times log_k(K) \times E)$, when the cluster number $k$ is fixed. $K$ is the number of $Blocks$, and $E$ is the dimension of tuple encodings.

Our proposed data reorganization method groups tuples that are either similar or share similar query access into the same $Block$. Nevertheless, due to the increased construction overhead, the acceptance and generalization of this method have been restricted. Actually, the performance bottleneck of this algorithm is associated with the HBC algorithm employed for clustering. Even though its time complexity has been reduced compared to traditional K-Means, it remains challenging for deployment in scenarios involving extensive data volumes. We will delve into the optimization of our method for large-scale data processing requirements further in Section~\ref{subsec:tp}.

\begin{algorithm}[htb]
	\color{black}
        \caption{ \textcolor{black}{Data Reorganization.}}
	\label{alg:dr}
        
	\begin{algorithmic}[1] %这个1 表示每一行都显示数字
		\Require
             Table $D$, Representative workload $W$
		\State $ tupleEncode(D,W) $;
            \State $ Blocks \gets \emptyset $;
            \State $ hbcTree \gets \emptyset $;
            \State $ queue \gets [D] $;
            \While{\textit{queue is not empty}} 		
            \State $ node \gets queue.pop()$;
            \State $ k \gets dynamicAdjust(K,node)$;
            \State $ children \gets balancedKmeans(node,k)$;
            \State $ treeUpdate(hbcTree,node,children)$
            \For{\textit{child in children}}
                \If{$child.size \le block\_size$}
                \State $ block \gets blockCreate(child)$
                \State \textit{add block to Blocks}
                \Else
                \State \textit{add child to queue}
                \EndIf
            \EndFor
            \EndWhile
            \State $hbcTree \gets hierarchicalStore(Blocks,hbcTree)$
		\\ 
		\Return $hbcTree$; %算法的返回值
	\end{algorithmic}
        \color{black}
        
\end{algorithm}

\vspace{-0.5em}
\subsection{Table Partition} \label{subsec:tp}

When dealing with algorithms characterized by high computational complexity, the notion of accelerating them through parallelization or distributed methods naturally arises as a means to enhance their applicability in large-scale data scenarios. This notion is particularly pertinent to the data reorganization algorithm introduced in Section~\ref{subsec:dr}, which revolves around the management of input data scale, specifically, the original table. 
To effectively control the data volume undergoing the execution of Algorithm~\ref{alg:dr}, we formulate our table partition algorithm in Algorithm~\ref{alg:tp}. The algorithm encompasses four key components, i.e., Pre-Segmentation (Line 1), Workload-aware Page Encoding and Filtering (Lines 2-4), Hot Pages Clustering (Line 5), and Cold Pages Reorganization (Lines 6-8). The output of this algorithm comprises a partition scheme for hot data and a query routing tree for cold data. The key components are described as follows. 

\textbf{Pre-Segmentation}: Ensuring the effectiveness of table partitioning, we increase the processing granularity from the tuple level to the page level. To achieve this goal, we employ a K-D Tree for pre-segmenting the original table into pages. A K-D Tree is a tree-like data structure used to partition data points in a K-dimensional space for efficient range queries and nearest neighbor searches. We implement a version that selects the data dimension with the highest variance as the splitting dimension and splits from the median. The K-D Tree iteratively partitions the high-dimensional data space until each subspace contains fewer than $2 \times Page\_size$ data tuples. We choose the K-D Tree to partition the table into data pages for three reasons: (1) easy to construct, (2) efficiently support fast query retrieval requirements, (3) data partitioning follows the principle of locality, which allows for the aggregation of tuples with similar domains within the same or neighboring pages.
For the scenarios without strict time constraints on block generation, tables can also be partitioned into pages using workload-aware methods such as Qd-tree~\cite{Yang2020QdtreeLD}.

\textbf{Workload-aware Page Encoding}: \textbf{BG} employs the K-D Tree to expeditiously direct representative workloads $W$ to various data pages and generates feature vectors for these pages with respect to the workloads. Specifically, a 0-1 vector of $|W|$ dimensions is maintained for each data page, with the $i$th position set to 1 for pages accessed by the $i$th query, and 0 otherwise.

\textbf{Page Filtering}: Following the encoding of pages, we tally the occurrences of '1's in the feature vector of each page, indicating the query access frequency to that specific page. These counts are then sorted.
Filtering cold data pages holds significance in diminishing the data volume subjected to the data reorganization algorithm for restructuring. 
Moreover, an ancillary advantage of this approach lies in its capacity to enhance clustering performance. This is rooted in the observation that the feature vectors tailored for cold data, designed in alignment with the workload, inherently exhibit sparsity. Consequently, conventional clustering algorithms like K-Means, which rely on Euclidean distance, are ill-suited for clustering such feature vectors.

\textbf{BG} supports two different methods for filtering cold pages according to filter criteria. The first method filters pages with access frequencies falling below a given threshold, referred to as \textit{Hard Filtering}. The second method, \textit{Soft Filtering}, retains additional cold pages situated at the edge of the hot zones. Grounded in the principle of locality, this approach ensures substantial robustness, even in the face of slight fluctuations in future workloads. Below, we will explain \textit{Soft Filtering} in detail.

\textit{Soft Filtering}: In the preceding two steps, we partition the original table into data pages and map queries based on Max-Min index to the corresponding pages to build feature vectors for each page. At this stage, we maintain two additional sets of statistics for each page. Specifically, for data page $p_{i}$, we record (1) the positional feature, i.e., the mean of the Max-Min index corresponding to $p_{i}$, denoted as $l_{i}$, and (2) the heat feature, i.e., the number of queries accessed, denoted as $h_{i}$.

Subsequently, we arrange the data pages along a space-filling curve~\cite{Hilbert1935DritterBA} based on their position feature, resulting in the sequence of data page positions $P=[p_1,p_2,\ldots,p_K]$. The space-filling curve serves as an effective method for dimensionality reduction in high-dimensional data. In our system, we opted for the Hilbert curve due to its optimal performance in sorting pages~\cite{Hilbert1935DritterBA}. This transformation converts the challenge of filtering and segregating hot and cold pages into a sequence partitioning problem, where each subsequence corresponds to a spatially contiguous region.

We segment the sequence based on the popularity feature of pages to distinguish between hot zones and cold zones. This is accomplished through a two-round scanning. In the first round, pages with access frequencies below the popularity threshold are labeled as cold pages, and vice versa. Simultaneously, consecutive cold pages (hot pages) are merged into a cold zone (hot zone). In the second round, if the size of a cold zone between two hot zones is smaller than the size threshold, we merge these three segments into a larger hot zone. As a result, pages are partitioned into distinct hot zones and cold zones.

\textbf{Hot Pages Clustering}: The clustering approach for hot pages, associated with both \textit{Hard Filtering} and \textit{Soft Filtering}, exhibits subtle distinctions. In the case of \textit{Hard Filtering}, \textbf{BG} conducts clustering on the unfiltered pages based on their feature vectors, with each cluster corresponding to a sub-table in the original table. In the context of \textit{Soft Filtering}, considering its effective separation of hot zones, the objective of employing clustering techniques is to break down excessively large hot zones into several relatively smaller ones. Here we apply the hierarchical iterative clustering method in Section~\ref{subsec:dr}. This method is designed to prevent the formation of excessively large sub-tables, as uneven partition might compromise the efficacy of parallel acceleration strategies.

\textbf{Cold Pages Reorganization}: To meet the demands of cold data management, we aggregate the filtered-out cold data pages into a table. Following this, we employ the K-D Tree to structure this table into $Blocks$, which are subsequently positioned in the cloud for centralized management. The K-D Tree, serving as an effective indexing structure for cold data, is preserved and backed up by the \textbf{BG} module for both edge and terminal devices. It works in conjunction with the hierarchical clustering trees of individual sub-tables, contributing to the overall efficiency of query routing and data localization.

\begin{algorithm}[htb]
	\color{black}
        \caption{ \textcolor{black}{Table Partitioning.}}
	\label{alg:tp}
        
	\begin{algorithmic}[1] %这个1 表示每一行都显示数字
		\Require
             Table $D$, Represent workload $W$
		\State $ Pages,routeTree \gets KDTree(D, page\_size) $;
            \State $ queryRoute(W,Pages,routeTree$;
            \State $ pageEncode(Pages)$;
            \State \small{$ hotPages,coldPages \gets pageFilter(Pages,freq\_limit)$};		
            \State $ hotZones \gets HBC(hotPages)$;
            \State $ coldZones \gets pageMerge(coldPages)$;
            \State \small{$ Blocks,indexTree \gets KDTree(coldZones,Block\_size)$};
            \State $ cloudStore(Blocks)$
		\\ 
		\Return $hotZones,indexTree$; %算法的返回值
	\end{algorithmic}
        \color{black}
        
\end{algorithm}
\vspace{-0.5em}

\section{Data Migration and Scheduling} \label{sec:dm}

In this section, We introduce \textbf{Brame}'s compact three-tier storage scheduling approach for \textbf{CEDC}, using $Blocks$ as the basic unit. The practical requirements of this architecture are detailed in Section~\ref{subsec:pr} and the scheduling workflow is outlined in Section~\ref{subsec:ss}. The data migration algorithm for cloud-edge periodicity and the real-time cache replacement strategy between cloud-edge and end are explained in Sections~\ref{subsec:dsce} and Section~\ref{subsec:crcee}, respectively.

\vspace{-0.5em}
\subsection{Practical Requirements} \label{subsec:pr}

Our goal for \textbf{CEDC} is to minimize the frequency of communication and data interactions between terminal devices and the cloud. To achieve this goal, \textbf{Brame} must efficiently identify hot data, which are frequently requested by users, and place it closer to users on edge or end devices. Furthermore, given the dynamic nature of user requirements, \textbf{Brame} should be proficient in scheduling and migrating data between different device levels in an online fashion. After consulting with professionals, we have discerned two distinct user data demand categories: common and personalized requirements. Hence, \textbf{Brame} is required to employ distinctive data storage strategies for these diverse demand patterns. For instance, when considering a group of employees within the same business line, common data demands reflect the inherent characteristics of the business, while personalized demands depend on individual employee roles.

\vspace{-0.5em}
\subsection{Scheduling System} \label{subsec:ss}

To meet the diverse requirements of \textbf{CEDC}, we design \textbf{Brame} to be an online hierarchical data storage and scheduling framework. Our underlying rationale is that $Blocks$ that have been frequently accessed in recent times are likely to remain important in the near future. To reduce cloud-end interactions, these significant $Blocks$ should be located at the edge or end. We employ a "temperature" metric to quantitatively measure $Blocks$ significance, which reflects data access frequency and timeliness. \textbf{Brame} assigns distinct storage functions to cloud, edge, and terminal devices, respectively. Within our design, $Blocks$ with low temperature are situated in the cloud, while hot $Blocks$ that are frequently requested by a group of users find their place on edge servers. Individual users' personalized hot $Blocks$ are cached on terminal devices. The K-D Tree, as expounded in Section~\ref{sec:bg}, and the hierarchical clustering forest, constructed from hierarchical clustering trees of various sub-tables, acts as the index for cold and hot data, respectively. These structures are cached on edge and terminal devices, facilitating efficient query routing and expedited data localization.

In implementations, the query router combines index-based query routing with a strategy of directly scanning leaves. For small leaf sets, we directly scan the leaves to determine the $Blocks$ to be accessed. With larger leaf sets, we first scan the index tree to a certain depth and then scan the leaves corresponding to the intermediate nodes that satisfy the query, skipping $Blocks$ that do not meet the filter conditions.

To meet the varied demands of \textbf{CEDC} storage, \textbf{Brame} has been equipped with appropriate data migration strategies. In particular, \textbf{Brame} periodically aggregates the user-shared data requirements. Afterwards, \textbf{Brame} regularly triggers temperature calculation and data migration modules to ascertain the allocation plan for $Blocks$ between the cloud and the edge. For personalized data needs, \textbf{Brame} delegates migration task to cache replacement strategies like CLOCK~\cite{Jiang2005CLOCKProAE}~\cite{Lee2015MCLOCKMP} on terminal devices.

\vspace{-0.5em}
\subsection{Data Scheduling between Cloud and Edge} \label{subsec:dsce}
In this section, we delve into the technical details of periodic data migration between the cloud and edge, including assessing the significance of each $Block$ in the current timeframe based on a temperature model. We also model the problem of hierarchical placement of data between the cloud and edge as a \textit{0-1 knapsack problem} to determine the scheduling scheme for data migration.

\textbf{Temperature Model}: We employ a temperature model to provide a quantitative characterization of $Block$ significance. In contrast to methods like LRU, which assess data's heat based on the relative position of data within a specific data structure, the temperature model offers the advantage of persistence and consistency. It retains the heat indication even when data transitions from a specific data structure, such as migrating from the edge to the cloud. Temperature is considered an inherent attribute of $Blocks$, and uniform rules are applied to calculate and update temperatures for $Blocks$ within the same system, ensuring consistent management at both the edge and the cloud. A robust temperature model, in our view, should capture the characteristics of data access frequency and timeliness, where $Blocks$ that have been frequently accessed in recent times should exhibit higher temperatures. Moreover, the influence of historical data on the current temperature of a $Block$ diminishes over time. Inspired by Siberia~\cite{Levandoski2013IdentifyingHA}, we employ exponential smoothing to model $Block's$ temperature, effectively reconciling these two requirements and yielding favorable experimental results.

\textit{Heat Model}: For $Block$ $b$, let its temperature at time $t$ be $H(t)$. After a time interval $\Delta t$, its temperature becomes:
\begin{equation}
H(t + \Delta t)=\gamma \times H(t) + (1 - \gamma) \times hits
\end{equation}

Here, $hits$ is the frequency of accesses within the time period $\Delta t$, $\gamma$ is the decay coefficient, and the first term models the influence of historical temperature on the current temperature, while the second term reflects the recent data access intensity.

\parindent=0pt
\textbf{Data Placement}: We model the data placement between the cloud and the edge as a \textit{0-1 knapsack problem}~\cite{Du2011DesignAA}.
\parindent=9pt

\textit{Binary Knapsack Problem}: Given $2n+1$ positive integers $S$, $s_1,s_2,\ldots,s_n$ and $c_1,c_2,\ldots,c_m$, solve the (0-1) integer programming problem:
\[Maximize\quad c(x) = c_{1}x_{1} + c_{2}x_{2} + \ldots + c_{n}x_{n},\]
\[subject\quad to\quad s_{1}x_{1} + s_{2}x_{2} + \ldots + s_{n}x_{n} \le S,\quad where\]
\[ x_1,x_2,\ldots,x_n\in \{0,1\}\]
In our research context, the storage budget of the edge device is $S$, the volume of $Block$ $i$ is $s_i$, which signifies the number of tuples contained within it, and the value $c_i$ corresponds to the current temperature of $i$. If $i$ is stored at the edge, the corresponding $x_i$ is set to 1. 

The knapsack problem can be solved using various solver techniques: dynamic programming algorithm can find the optimal solution to the knapsack problem with a time complexity of $O(n^{3}Mlog(MS))$, where $M=max\{c_k|1\le k \le n\}$, making it a pseudo-polynomial time algorithm. A greedy algorithm with an approximation ratio of 2 can find a solution in linear time. In practical applications, the choice of solver can be based on time budget and precision requirements.

By integrating the temperature model with the 0-1 knapsack problem solver, we establish a fundamental model for data migration and scheduling. This model operates in periodic cycles, where it is awakened and invoked. We define the time interval from completing data migration to the next activation of the temperature calculation module as a time cycle. 

Algorithm~\ref{alg:bm} outlines the process of the cloud-edge block migration algorithm over a specific time period. The initial line represents the data preparation stage. Lines 2-4 execute queries and update the statistical information for each $Block$. Lines 5-6 compute the temperature of each $Block$, considering current query hit patterns. Line 7 utilizes the 0-1 knapsack solver to establish the allocation plan for each $Block$. In line 8, the information concerning the storage location of $Blocks$ in the K-D Tree and hierarchical clustering forest is updated. The time complexity for the temperature calculation and update is $O(|W|\times n)$, while the complexity for data migration plan generation is $O(n\times log(n\times M\times S))$, where $|W|$ is the size of the workload batch, and $n$ is the number of $Blocks$.

\begin{algorithm}[htb]
	\color{black}
        \caption{ \textcolor{black}{Block Migration.}}
	\label{alg:bm}
        
	\begin{algorithmic}[1] %这个1 表示每一行都显示数字
		\Require
             A set of queries intercepted within a given time interval $Batch$, Query routing structure for cold data $kdTree$, Query routing structure for hot data $hbcForest$
		\State $ Blocks \gets getLeaves(kdTree, hbcForest) $;
        \For{\textit{Q in Batch}}
            \State $ accessBlocks \gets queryRoute(Q,kdTree,hbcForest)$;
            \State $ hitsUpdate(Blocks,accessBlocks)$;
        \EndFor
        \For{\textit{Block in Blocks}}
            \State $ Block.heat \gets heatCompute(Block.hits)$;
        \EndFor
        \State $ Blocks \gets blockPlacer(Blocks)$;
        \State $ kdTree,hbcForest \gets placeUpdate(Blocks)$;
            
		\\ 
		\Return $kdTree,hbcForest$; %算法的返回值
	\end{algorithmic}
        \color{black}
        
\end{algorithm}

\vspace{-0.5em}
\subsection{Data Migration between Cloud / Edge and End} \label{subsec:crcee}

The data migration between the cloud and the edge is designed to capture the collective requirements of a group of users over a specific period. In contrast, data scheduling involving the cloud, edge, and end-users encompasses nuanced and highly individualized data needs. Our goal is to deliver tailored data services to end-users. The basic idea of our solution is to configure the end cache based on the behavior of end-users. Unlike the periodic execution of data migration from the cloud to the edge, real-time data scheduling at the end is imperative due to the dynamically changing data requirements of users. In light of these considerations, \textbf{Brame} entrusts the responsibility of data migration between the cloud \ edge and end to the cache replacement strategy on the end.

Specifically, the system actively monitors query access requests initiated by end-users. It subsequently identifies the necessary $Blocks$ and their locations by examining the query routing structures deployed at both the edge and end. In case of a cache miss, meaning the required $Block$ is absent at the end, the terminal device requests the $Block$ from the corresponding cloud or edge device. Upon obtaining the required $Block$, the terminal device reconfigures its cached $Blocks$ with a user-selected cache replacement strategy like LRU.

\section{Experiments} \label{sec:exp}

In this section, we conduct extensive experiments to evaluate the performance of \textbf{Brame}. 

\vspace{-0.5em}
\subsection{Experimental Settings} \label{subsec:set}

\textbf{Platform}: 
We use Python 3.8 to build a simple prototype system on a server with 12 Intel(R) Xen(R) Silver 4210R CPUs and 520 GB RAM to simulate the scenario of \textbf{CEDC}. To guarantee the rationality and authenticity of the experimental results, we choose platform-independent metrics to evaluate the performance of our approach.

\parindent=0pt
\textbf{Baseline Methods}: We selected the following baseline methods for comparison:
\parindent=9pt

(1) Key-Order: Sorting tuples in the table lexicographically, with every $Block\_size$ tuples encapsulated into a $Block$.

(2) K-D Tree: For a K-D Tree, we implemented the standard version that selects the splitting dimension through polling and splits the data based on the median. We generate $Blocks$ based on the leaves of the tree.

(3) Curve: Reordering tuples in the table along the Hilbert curve, with every $Block\_size$ tuples forming a $Block$.

\parindent=0pt
\textbf{Dataset}: We choose two real datasets with different scales and features for experimental evaluation, as detailed in Table.~\ref{TABD}:
\parindent=9pt

\begin{table}[thbp]
\caption{Datasets}
\vspace{-1.0em}
\label{TABD}
\centering
\begin{tabular}{|c|cccc|}
\hline
Dataset        & Size(MB)   & Cols/Cat  & Rows  &  Domain   \\ \hline
Power~\cite{power}    & 110.8   & 7/0     & 2.1M    & $10^{17}$          \\ \hline
DMV~\cite{dmv}    & 972.8   & 11/10   & 11.6M    & $10^{15}$           \\ \hline
\end{tabular}
\vspace{-0.5em}
\end{table}

We preprocess all the data in the same way, i.e., Max-Min normalization for numerical columns, and for non-numerical columns, extracting unique values, sorting them, mapping them to corresponding positive integer values, and then normalizing them.

\parindent=0pt
\textbf{Workload}: We adopt the method similar to~\cite{Wang2020AreWR} to build a unified workload generator for query generation. Considering that this paper focuses on the single-table data storage and scheduling problem, we only evaluated read-only queries in the following form:
\textit{Select  $*$  From T Where  $l_1 < col_1 < u_1$ and \ldots and  $l_k < col_k < u_k$}
\parindent=9pt

To simulate real scenario, we generate workloads as follows. First, we use the workload generator mentioned above to generate a representative workload. Then, we match the workload with randomly generated query arrival rate curves, which is a time series where each value represents the frequency of the query within a specified time interval. Based on the representative queries and their corresponding query arrival rate curves, we construct training and testing workloads. In the generation, we allow the queries to have a slight skew compared to the original query. Finally, we aggregate queries within each time interval to build the final training and testing workloads.

\parindent=0pt
\textbf{Evaluation Metrics}: We use the following metrics for performance evaluation.
\parindent=9pt

(1) \textit{Tuple-level query hit rate} (\textit{THR}): For a query $Q$, let $B_Q$ be the set of $Blocks$ to access, and let $B_C$ be the set of $Blocks$ in the system's cache at the current time. Let $A(Q,B)$ represent the tuples in $Block$ $B$ that satisfy $Q$. We define $THR(Q)=\frac{\sum_{B\in B_Q\cap B_C}|A(Q,B)|}{\sum_{B'\in B_Q}|A(Q,B)|}$. For cloud-edge data migration, we calculate the ratio of the number of query-satisfied tuples to the number of required tuples among the tuples stored at the edge. For cloud-edge-device cache replacement, \textit{THR} is the cache hit rate of the end device at the current time. \textit{THR} is the most direct metric to measure the effectiveness of data migration strategies and is widely used.

(2) \textit{Block-level query hit rate} (\textit{BHR}): We define $BHR(Q)=\frac{\sum_{B\in B_Q\cap B_C}1}{\sum_{B'\in B_Q}1}$. For cloud-edge data migration, this is the ratio of the number of $Blocks$ placed on the edge side that satisfies $Q$ to the total number of $Blocks$ requested. For cloud-edge-device cache replacement, this is the ratio of the number of $Blocks$ in the end device's cache pool that satisfy the query to the total number of $Blocks$ need to be scanned. Since \textbf{Brame} manages data in units of $Blocks$, we design the \textit{BHR} metric based on \textit{THR} for performance evaluation.

(3) \textit{Partition construction Time}: We use this metric to evaluate the construction cost of each method.

\parindent=0pt
\textbf{Parameter Settings}: For the Power dataset, we set the $Block\_size$ to 2048, and for the DMV dataset, we set the $Block\_size$ to 8192. We conduct ablation experiments on $Block\_size$ in Section~\ref{subsec:ve}. For the cloud-edge data migration, we set the decay coefficient of the temperature model $\gamma = 0.6$. For cloud-edge-device data scheduling, we choose LRU as the cache replacement strategy, and we conduct comparative experiments on the cache replacement strategy in Section~\ref{subsec:ve}.
\parindent=9pt

\vspace{-0.5em}
\subsection{Feasibility Experiments} \label{subsec:mvf}

In this section, we analyze and demonstrate the rationality and feasibility of taking $Block$ as the basic unit of storage management. Considering that the minimum granularity of cache replacement in DBMS systems is $Page$ (Postgres), and the $Block$ proposed by us is actually a generalization of the concept of Page. It means that in real database cache management, the basic unit of data scheduling is $Block$ rather than tuple. This fact indicates the rationality and feasibility of taking $Block$ as the unit of data scheduling in cache replacement. Below, we experimentally show the benefits of taking $Block$ as the basic unit of data migration scheduling in cloud-edge data migration.

\begin{table}[thbp]
\caption{Tuple Hits Ratio, on Power}
\vspace{-1.0em}
\label{TABM}
\tabcolsep=2.1pt
\centering
\begin{tabular}{|c|cccccc|}
\hline
\small Cache Budget        & \small Key-order   & \small K-D Tree  & Curve  & Brame-H  & Brame-S  &Tuple \\ \hline
4\%    & 0.114   & 0.126   & 0.082   & 0.179  & 0.239  & 0.184       \\ \hline
8\%    & 0.182   & 0.216   & 0.213   & 0.291  & 0.359  & 0.304       \\ \hline
16\%    & 0.334   & 0.413   & 0.445   & 0.497  & 0.585  & 0.513       \\ \hline
32\%    & 0.603   & 0.699   & 0.729    & 0.789  & 0.852  &0.814       \\ \hline
\end{tabular}
\vspace{-0.5em}
\end{table}

As shown in Table.~\ref{TABM}, the performance of data-aware approach is comprehensively worse than Tuple baseline, while the workload-aware approach performs much better than data-aware approach and also shows competitive results compared with Tuple baseline. The purpose of taking $Block$ as the basic unit of storage management is to sacrifice accuracy to achieve lower data management costs. Tuple baseline is actually a special case where $Block\_size$ equals to 1. Our further analysis of $Block\-size$ shows that for data-aware approach, smaller $Block\-size$ can benefit from finer storage granularity, but overly small $Block\-Size$ can make block-based storage management meaningless, so there is a trade-off between performance and management costs. 

The workload-aware approach can fully capture the similarity of query access patterns among tuples, so it shows a performance that is not worse than Tuple baseline. As for our approach combined with \textit{Soft Filtering} algorithm, it can make better use of data and query locality features by focusing on cold pages at the edge of hot zones, so it also achieve high robustness when the future workload distribution slightly fluctuates. This also explains why this approach shows better results than the baseline. The other reason why it outperforms baseline is the design of the temperature model. However, considering that this is not the focus of our research, we will defer the study of temperature model to our next work.

\vspace{-0.5em}
\subsection{Comparisons} \label{subsec:ce}

In this section, we comprehensively compare \textbf{Brame} with the data-aware block generation method on two real datasets across two different downstream tasks.

\begin{figure}[thb ]
        \vspace{-0.5em}
	\centering
	\subfigure[Tuple Hit Ratio]{
		\label{p2a}
		\includegraphics[width=0.45\columnwidth]{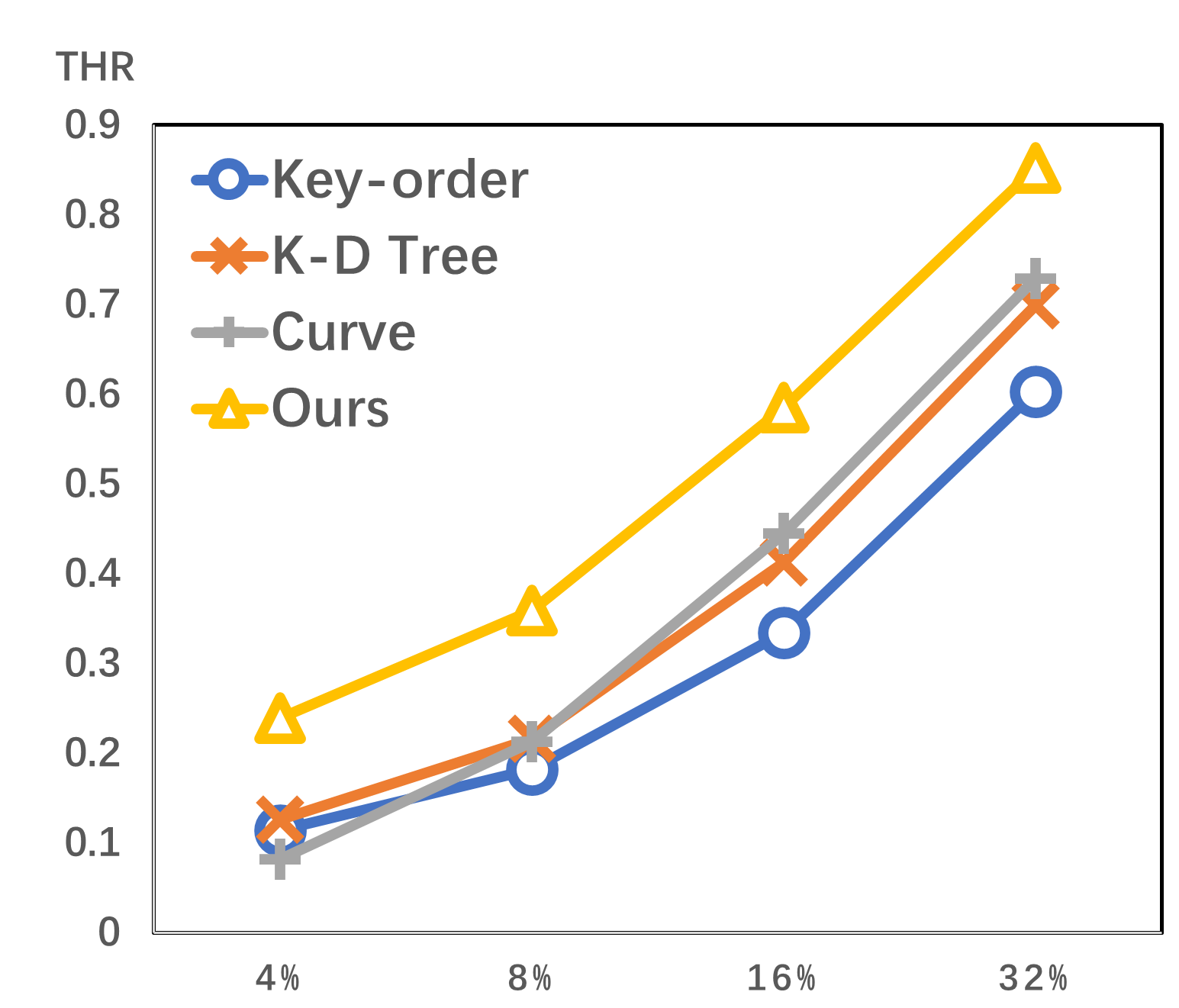 }}
	\subfigure[Block Hit Ratio]{
		\label{p2b}
		\includegraphics[width=0.45\columnwidth]{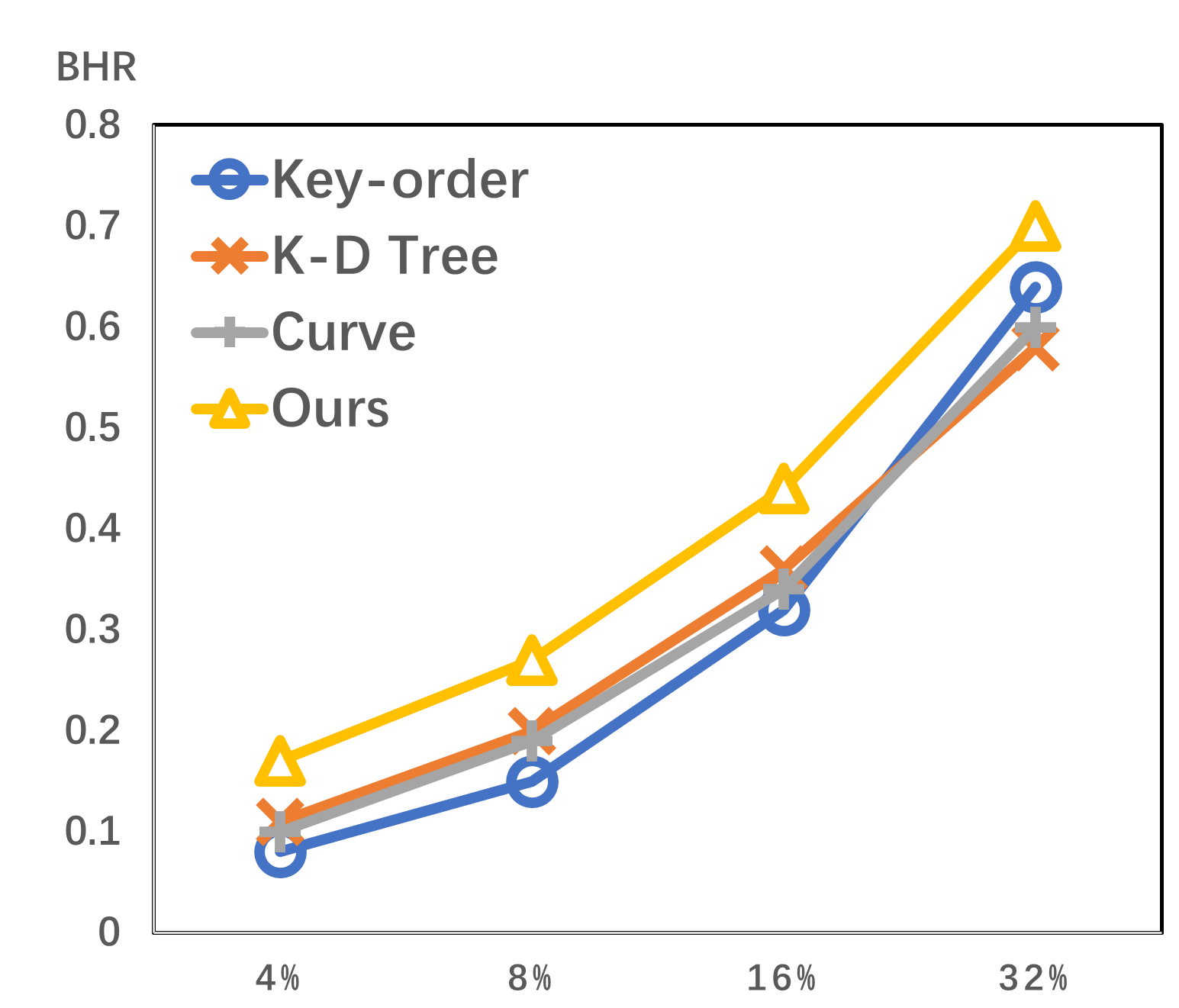 }}	
        \vspace{-0.5em}
	\caption{Experimental Results of Cloud-Edge Data Scheduling, on Power }  
         \label{p2}
         \vspace{-1.0em}
\end{figure}

\begin{figure}[thb ]
        \vspace{-0.5em}
	\centering
	\subfigure[Tuple Hit Ratio]{
		\label{d4a}
		\includegraphics[width=0.45\columnwidth]{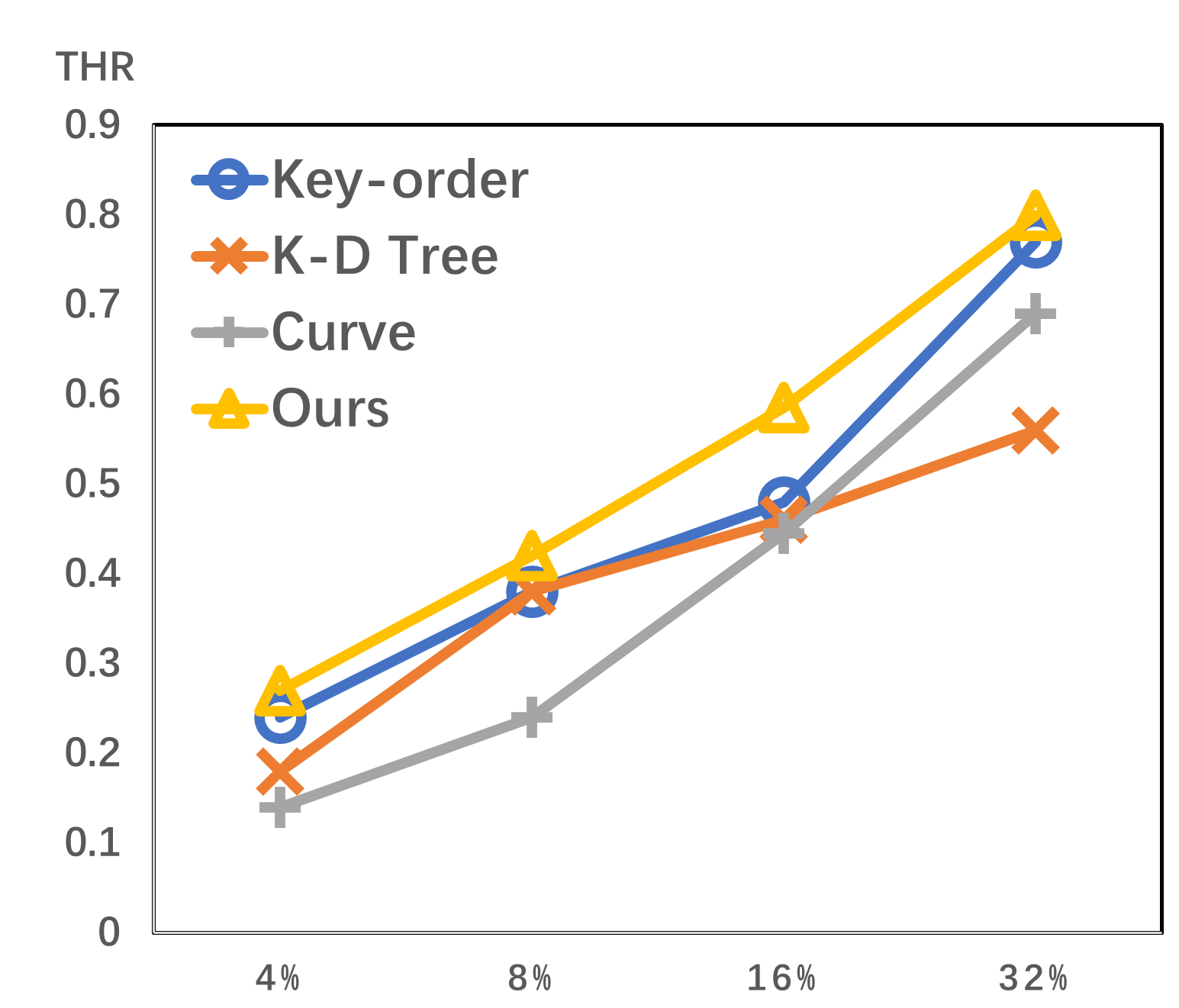 }}
	\subfigure[ Block Hit Ratio]{
		\label{d4b}
		\includegraphics[width=0.45\columnwidth]{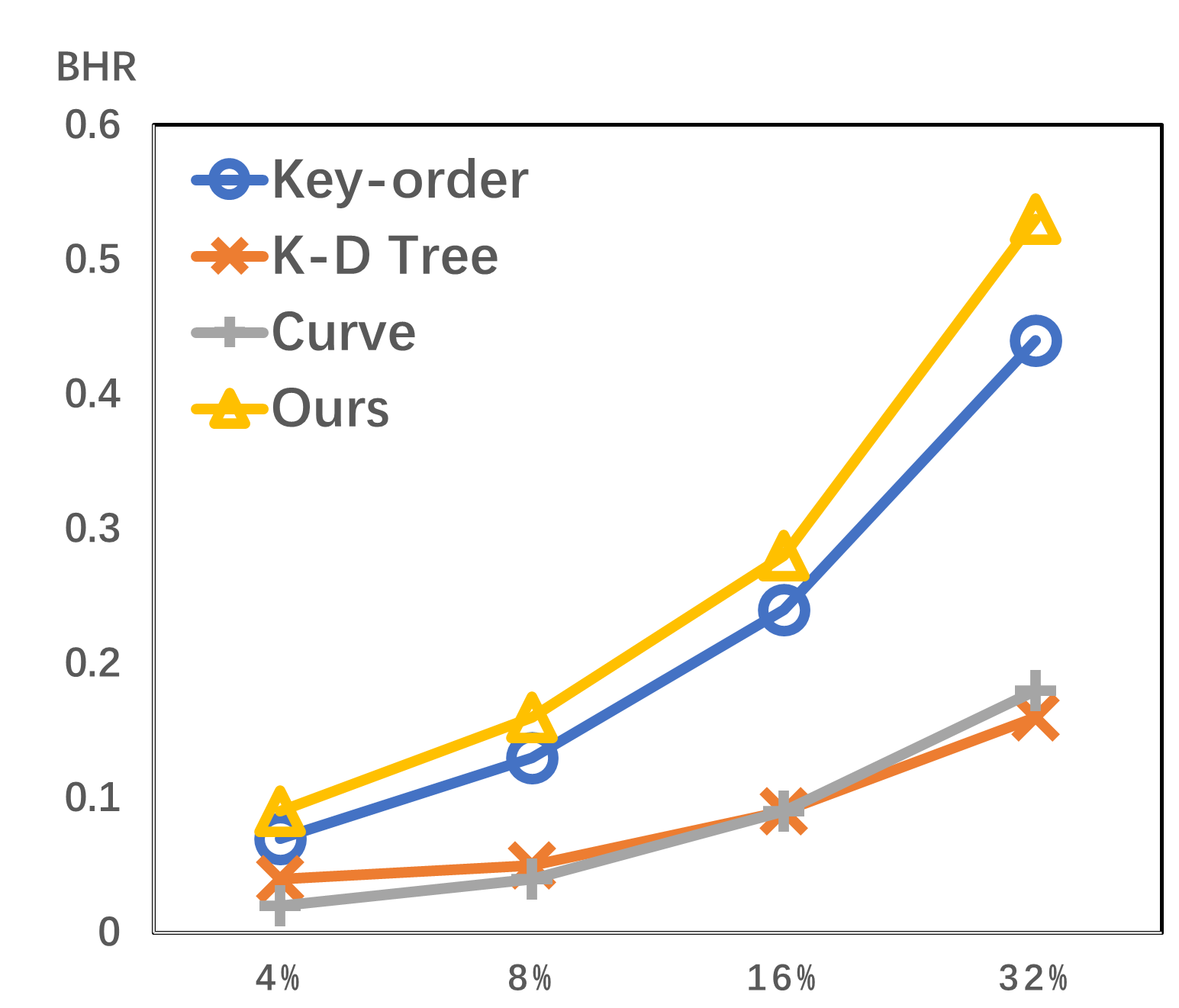 }}	
        \vspace{-0.5em}
	\caption{Experimental Results of Cloud-Edge Data Scheduling, on DMV }
         \label{d4}
        \vspace{-1.0em}
\end{figure}

\textbf{Evaluation on Cloud-Edge Data Migration}:
\autoref{p2} illustrates the performance of the four methods on the Power dataset. As the edge storage budget increases, the cache hit rates of all methods steadily rise. Under different cache budgets, \textbf{Brame} consistently outperforms the three baseline methods in terms of \textit{THR} and \textit{BHR}, exhibiting a performance improvement ranging from 0.5X to 1.0X. The reason lies in the inability of baseline methods to model and capture query access patterns among tuples. This results in tuples frequently accessed together being dispersed into different blocks. When storing and migrating data at the $Block$ level, this leads to storage redundancy (i.e., a $Block$ containing only a few tuples satisfying query needs), resulting in a waste of storage. \textbf{Brame} partially mitigates this by compactly organizing data with similar query access patterns. \autoref{d4} shows the performance of the four methods on the DMV dataset, showing results similar to \autoref{p2}. Notably, on the DMV dataset, the Key-order baseline outperforms the K-D Tree and Curve baselines, emphasizing the importance of analyzing dataset characteristics and distributions to determine an appropriate method for capturing data locality features. In comparison to Key-order, \textbf{Brame} not only captures data locality features but also considers the locality features of query access, resulting in a 10\% to 30\% performance improvement under both \textit{THR} and \textit{BHR} metrics.

\begin{figure}[thb ]
        \vspace{-0.5em}
	\centering
	\subfigure[ Tuple Hit Ratio]{
		\label{p3a}
		\includegraphics[width=0.45\columnwidth]{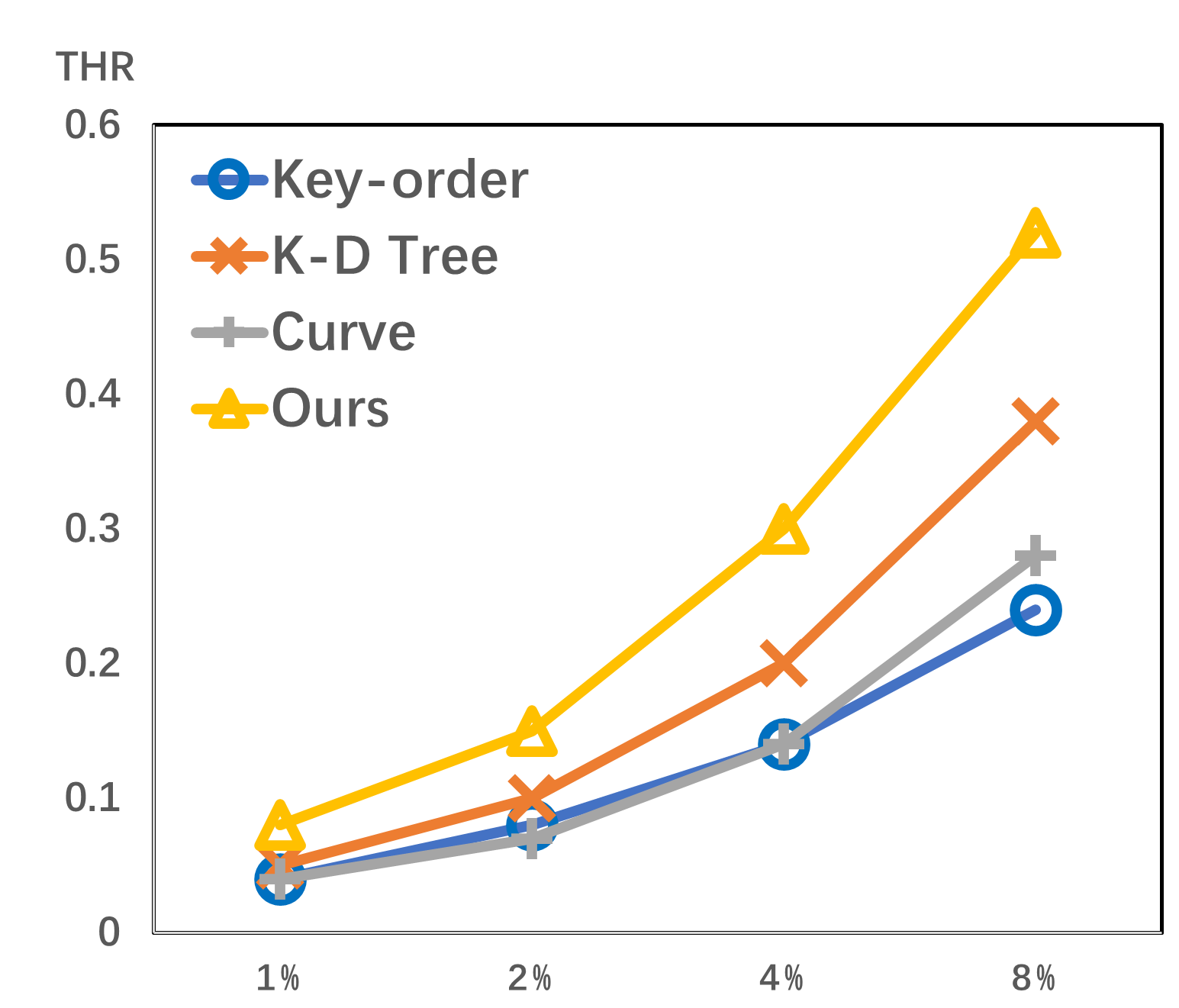 }}
	\subfigure[Block Hit Ratio]{
		\label{p3b}
		\includegraphics[width=0.45\columnwidth]{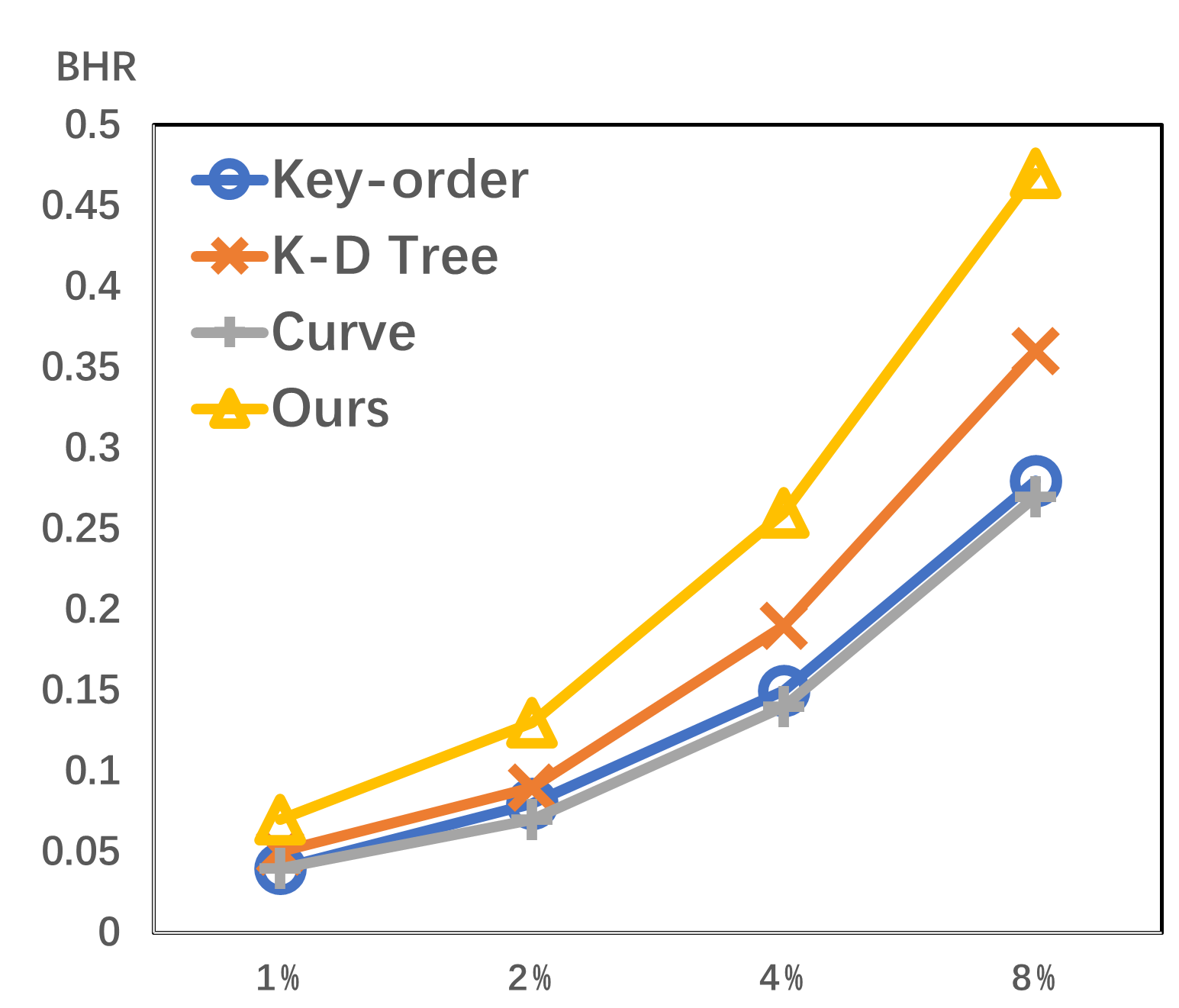 }}	
        \vspace{-0.5em}
	\caption{Exeprimental Results of Cloud-Edge-Device Cache Replacement, on Power }
         \label{p3}
         \vspace{-1.0em}
\end{figure}

\begin{figure}[thb ]
        \vspace{-0.5em}
	\centering
	\subfigure[ Tuple Hit Ratio]{
		\label{d5a}
		\includegraphics[width=0.45\columnwidth]{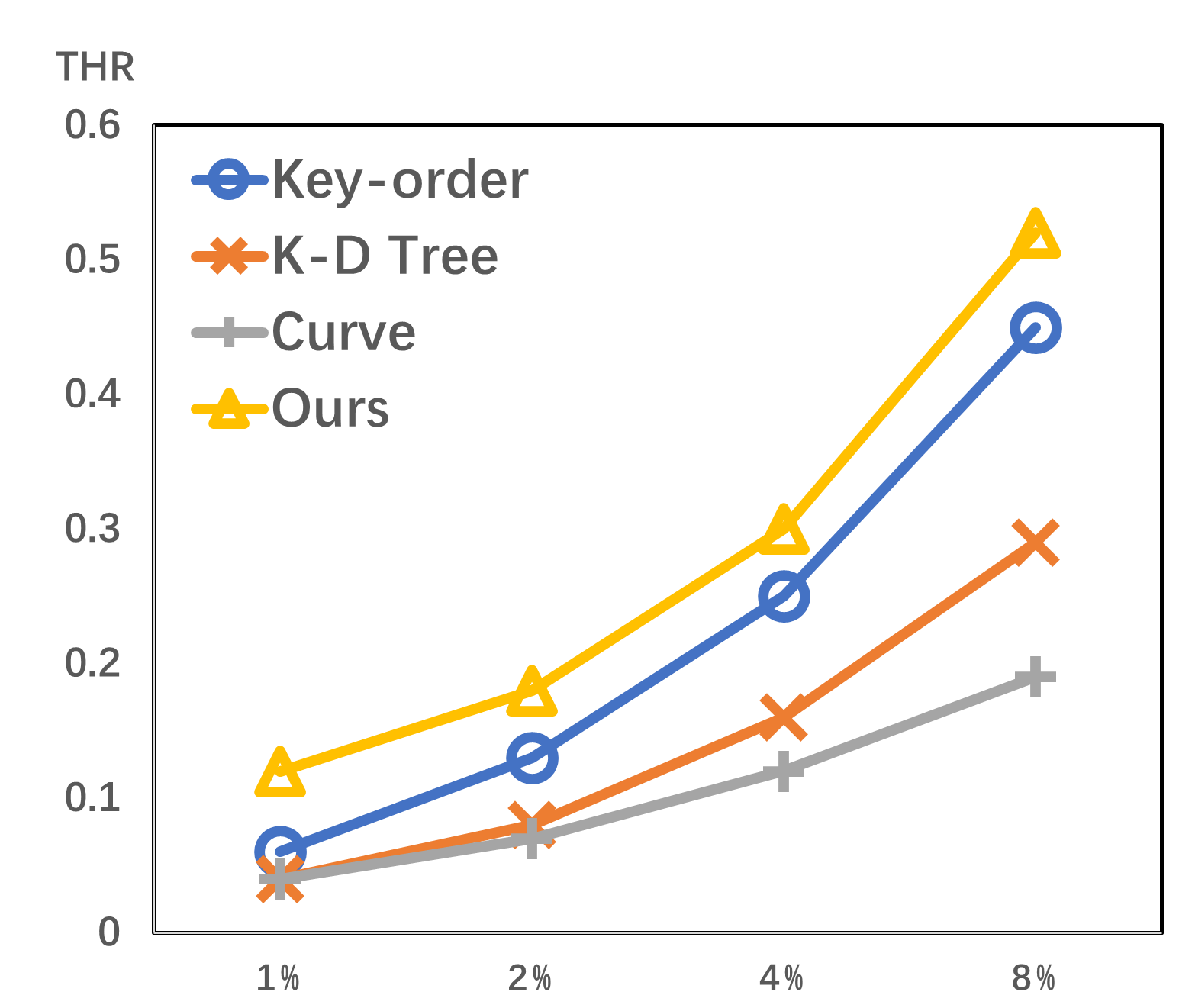 }}
	\subfigure[  Block Hit Ratio]{
		\label{d5b}
		\includegraphics[width=0.45\columnwidth]{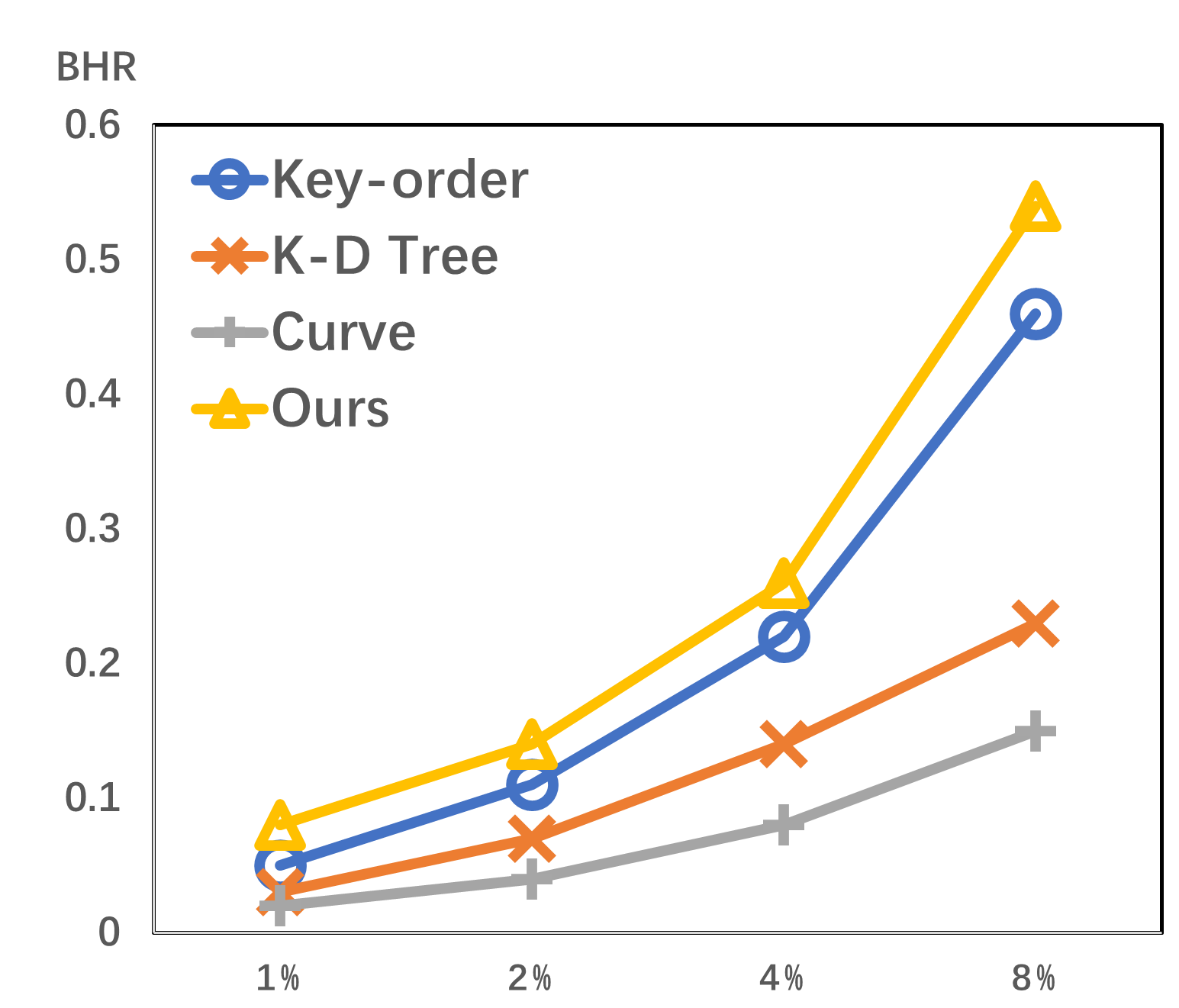 }}	
        \vspace{-0.5em}
	\caption{Experimental Results of Cloud-Edge-Device Cache Replacement, on DMV }
         \label{d5}
        \vspace{-1.0em}
\end{figure}

\textbf{Evaluation on Cloud-Edge-Device Cache Replacement}:
\autoref{p3} and \autoref{d5} depict the cache replacement task between cloud, edge and end devices. By changing the cache pool size of the end device and testing \textbf{Brame} on the Power and DMV datasets, we plot the average tuple cache hit rate curves and average block cache hit curves for the four different methods. Overall, \textbf{Brame} consistently outperforms various baseline methods. On the Power dataset, compared to the second-best K-D Tree baseline, \textbf{Brame} achieves a 0.3X to 0.7X improvement in both \textit{THR} and \textit{BHR} metrics. Similarly, on the DMV dataset, \textbf{Brame} leads by approximately 10\% to 30\% over the second-best Key-order baseline in \textit{THR} and \textit{BHR} metrics.

\textbf{Comparison of Block Generation Time Consumption}:
\autoref{TABC} reports the partition construction time for various methods on the Power and DMV datasets. The primary time expenditure of \textbf{Brame} lies in the Pre-Segmentation step within the Table Partition component and the separate execution of the HBC algorithm to reorganize data on each sub-table, accounting for 30\% and 60\% of the time, respectively. Further compression of the latter can be achieved by setting a larger $filter\_threshold$ hyperparameter in the page filter step of table partition.

\begin{table}[thbp]
\caption{Block Generation Time Consume }
\vspace{-1.0em}
\label{TABC}
\centering
\begin{tabular}{|c|cccc|}
\hline
Dataset        & Key-order   & K-D Tree  & Curve  &  Brame   \\ \hline
Power    & 3.12s   & 26.70s     & 130.67s    & 167.81s          \\ \hline
DMV    & 15.69s   & 146.40s   & 1139.82s    & 596.41s          \\ \hline
\end{tabular}
\vspace{-0.5em}
\end{table}

\vspace{-0.5em}
\subsection{Ablation Experiments} \label{subsec:ve}

In this section, we conduct ablation experiments on the essential components (cold-hot data page separation and filtering algorithms, cache replacement strategy) and key parameters ($Block\_size$) adopted in \textbf{Brame} to observe how they impact the overall performance. All experiments in this section are conducted using the Power dataset.

\begin{figure}[thb ]
        \vspace{-0.5em}
	\centering
	\subfigure[Tuple Hit Ratio]{
		\label{p6a}
		\includegraphics[width=0.45\columnwidth]{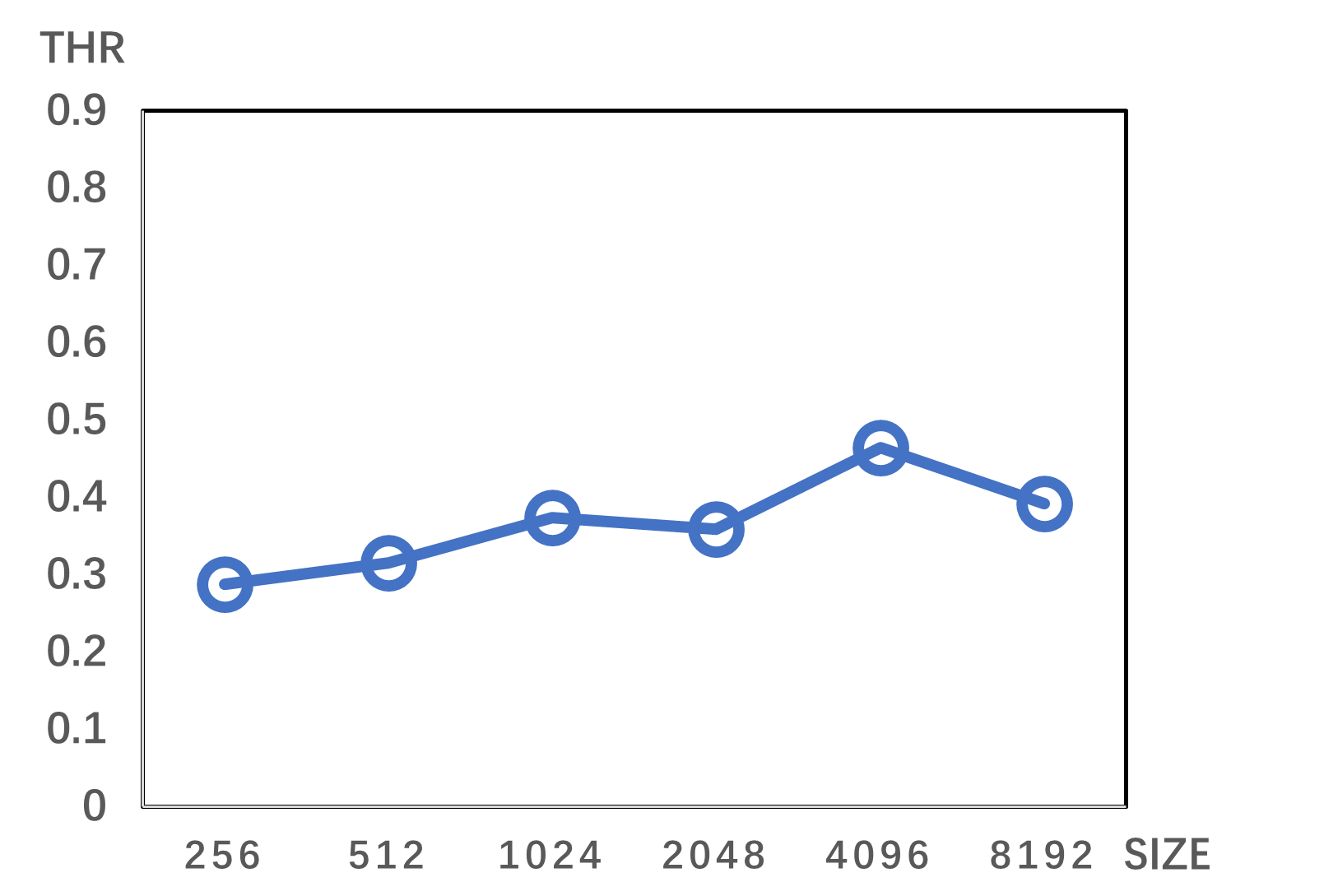 }}
	\subfigure[Block Hit Ratio]{
		\label{p6b}
		\includegraphics[width=0.45\columnwidth]{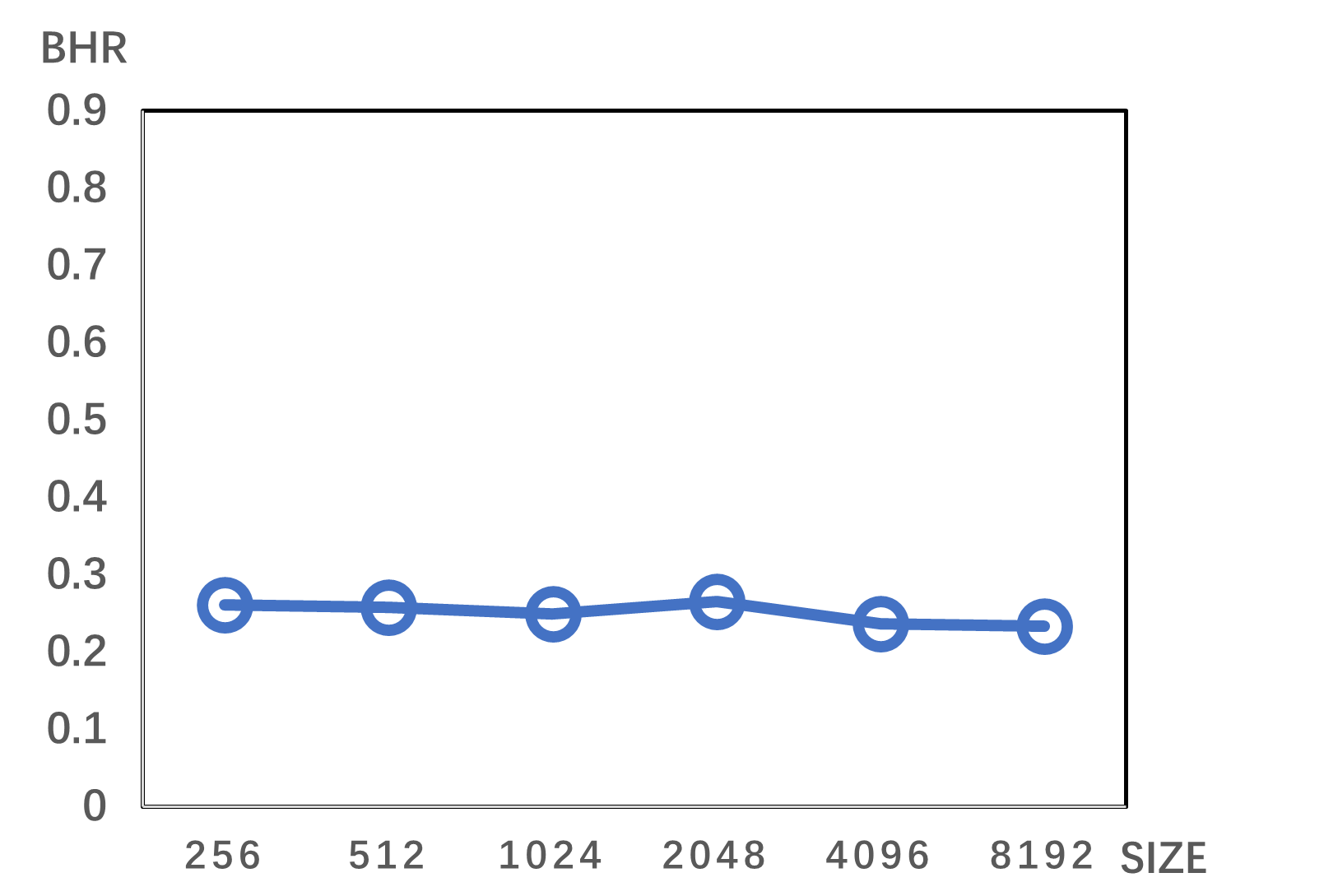 }}	
        \vspace{-0.5em}
	\caption{Sensitivity Study on $Block\_size$ in Cloud-Edge Data Scheduling}
	\vspace{-1.0em}
         \label{p6}
        
\end{figure}

\begin{figure}[thb ]
        \vspace{-0.5em}
	\centering
	\subfigure[Tuple Hit Ratio]{
		\label{p7a}
		\includegraphics[width=0.45\columnwidth]{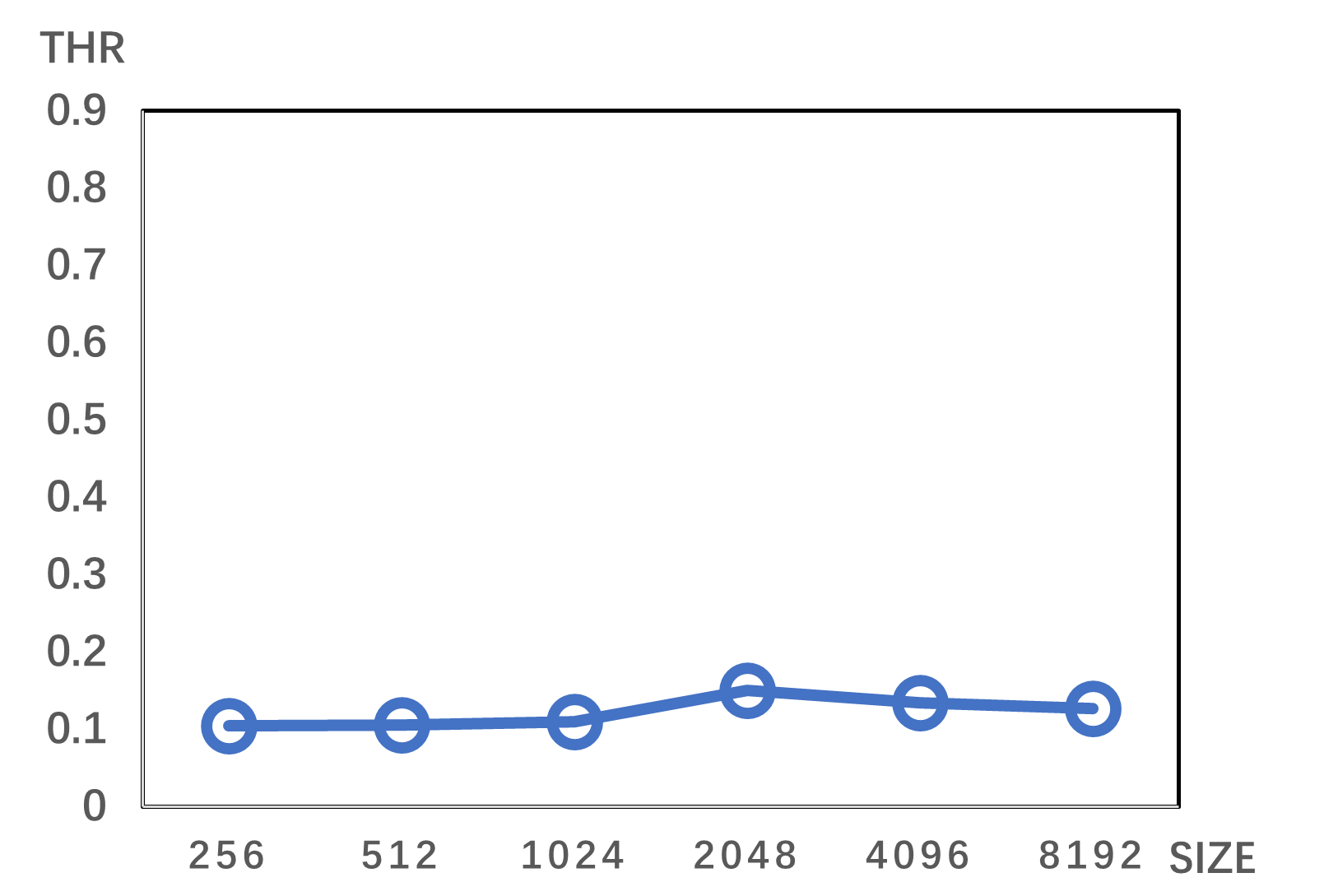 }}
	\subfigure[Block Hit Ratio]{
		\label{p7b}
		\includegraphics[width=0.45\columnwidth]{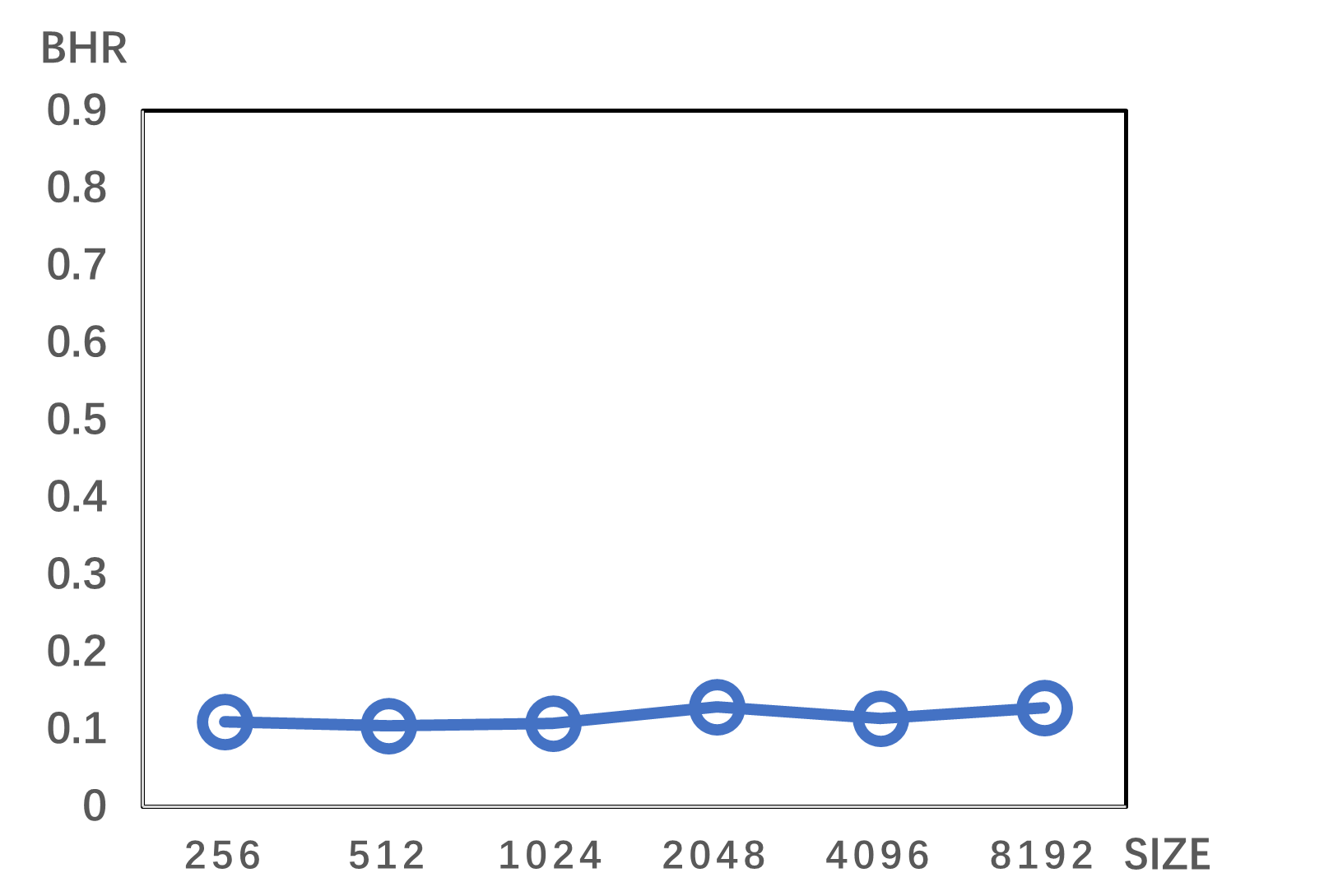 }}	
        \vspace{-0.5em}
	\caption{Sensitivity Study on $Block\_size$ in Cloud-Edge-Device Cache Replacement  }
        \vspace{-1.0em}
         \label{p7}
        
\end{figure}

\textbf{Effect of $Block\_size$}: In this part, we explore the impact of $Block\_size$ on the system performance. For the cloud-edge data migration task, where the edge cache budget is fixed at 8\% of the dataset's scale, as seen in \autoref{p6a} and \autoref{p6b}, we observe a general trend in which both the average tuple hit rate and the average block hit rate increase initially and then decrease with the increase in $Block\_size$. Similarly, for the cloud-edge-device cache replacement task, where the terminal device's buffer pool size is fixed at 2\% of the dataset size, the results are shown in \autoref{p7}, exhibiting a trend similar to the cloud-edge data migration task. We analyze that the reason for the initial increase and subsequent decrease in the average tuple hit rate is due to the HBC algorithm's execution, which allows the feature vectors of data tuples in intermediate nodes to be skewed and reorganized in a data-aware manner. When the $Block\_size$ is small, this situation occurs frequently, affecting the quality of block generation to some extent. Conversely, when $Block\_size$ is set too large, the precision of storage management will be sacrificed. Thus, it can be inferred that there is a trade-off between performance and management cost in setting $Block\_size$. The search and setting of the optimal $Block\_size$ are left for future research.

\textbf{Comparison of Page Filtering Strategy}: In this part, we evaluate the impact of page filtering strategies proposed in Section~\ref{subsec:tp} on \textbf{Brame}'s performance. \autoref{TABFA} and \autoref{TABFB} respectively show the performance of these two filtering techniques in the tasks of cloud-edge data migration and cloud-edge-device cache replacement. We find that combining $Soft$ $Filtering$ with different cache configurations yields better results across two evaluation metrics. We attribute this improvement to $Soft$ $Filtering$'s ability to retain cold pages adjacent to the hot zones. Considering that different data reorganization methods are applied to filtered and unfiltered pages in subsequent processing steps, it is nature to preserve pages with ambiguous cold-hot distinctions during the filtering step and perform workload-aware reorganization on their internal tuples. In the experiments, we observe that $Soft Filtering$ does not significantly increase the partition construction overhead.

\begin{table}[thbp]
\caption{$Soft$ $Filtering$ vs $Hard$ $Filtering$ in Cloud-Edge Data Migration}
\vspace{-1.0em}
\label{TABFA}
\tabcolsep=1.6pt
\centering
\begin{tabular}{|c|c|c|c|c|c|c|c|c|}
\hline
\small Cache Budget   &  \multicolumn{2}{c|}{4\%}  & \multicolumn{2}{c|}{8\%}  & \multicolumn{2}{c|}{16\%}  & \multicolumn{2}{c|}{32\%}  \\ \hline
\diagbox{\small{Method}}{\small{Metric}}    & THR   & BHR   & THR   & BHR  & THR   & BHR    & THR   & BHR   \\ \hline
Brame-H    & 0.179   & 0.123   & 0.291   & 0.211  & 0.497  & 0.375  & 0.789  & 0.619 \\ \hline
Brame-S    & 0.239   & 0.172   & 0.359   & 0.265  & 0.585  & 0.436  & 0.852  & 0.696  \\ \hline
\end{tabular}
\vspace{-0.5em}
\end{table}

\begin{table}[thbp]
\caption{$Soft$ $Filtering$ vs $Hard$ $Filtering$ in Cloud-Edge-Device Cache Replacement}
\vspace{-1.0em}
\label{TABFB}
\tabcolsep=1.6pt
\centering
\begin{tabular}{|c|c|c|c|c|c|c|c|c|}
\hline
\small Cache Budget   &  \multicolumn{2}{c|}{1\%}  & \multicolumn{2}{c|}{2\%}  & \multicolumn{2}{c|}{4\%}  & \multicolumn{2}{c|}{8\%}  \\ \hline
\diagbox{\small{Method}}{\small{Metric}}    & THR   & BHR   & THR   & BHR  & THR   & BHR    & THR   & BHR   \\ \hline
Brame-H    & 0.072   & 0.059   & 0.128   & 0.113  & 0.252  & 0.210  & 0.427  & 0.385 \\ \hline
Brame-S    & 0.082   & 0.070   & 0.150   & 0.129  & 0.297  & 0.258  & 0.516  & 0.472  \\ \hline
\end{tabular}
\vspace{-0.5em}
\end{table}

\begin{figure}[thb ]
        \vspace{-0.5em}
	\centering
	\subfigure[Tuple Hit Ratio]{
		\label{p8a}
		\includegraphics[width=0.45\columnwidth]{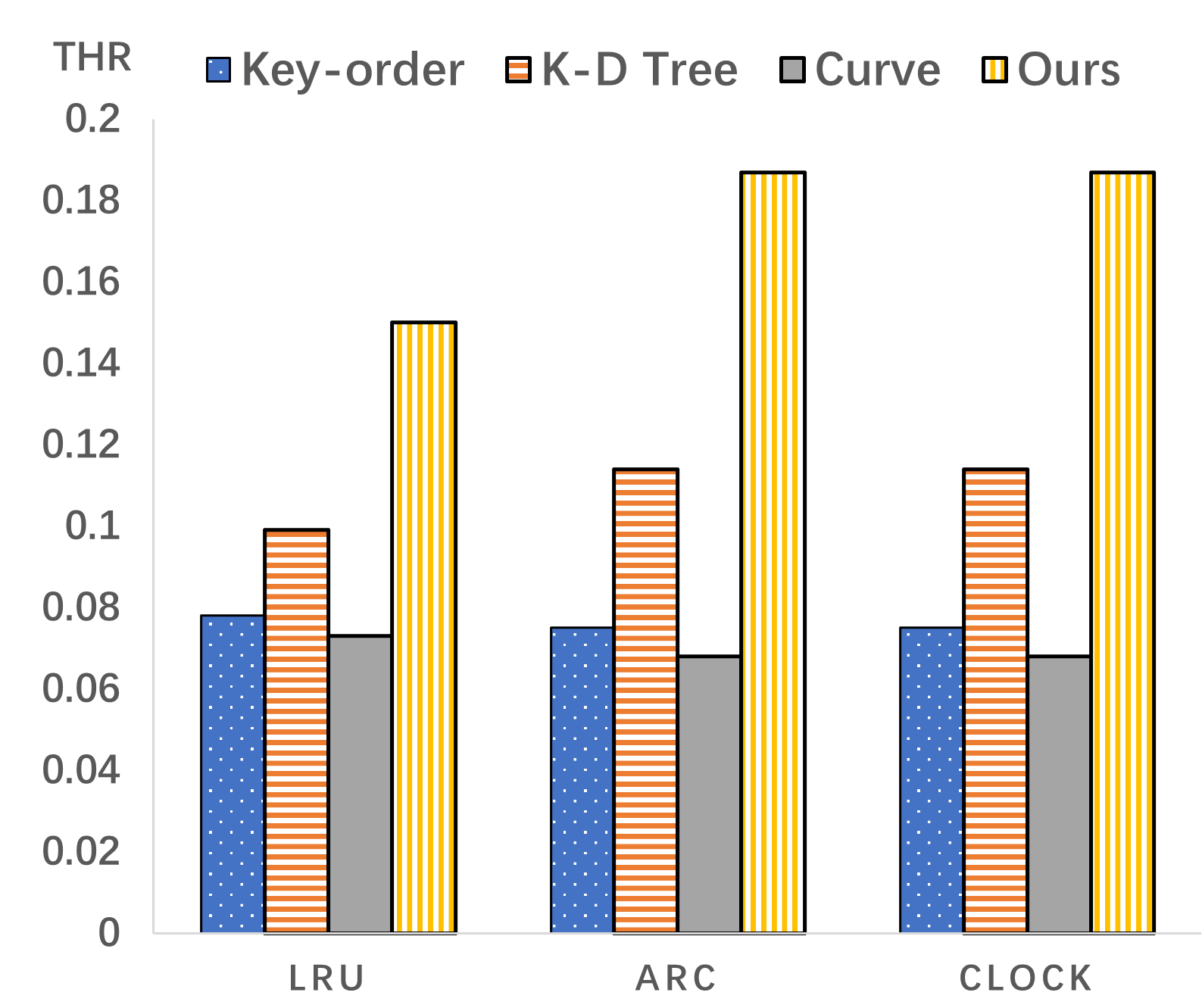 }}
	\subfigure[Block Hit Ratio]{
		\label{p8b}
		\includegraphics[width=0.45\columnwidth]{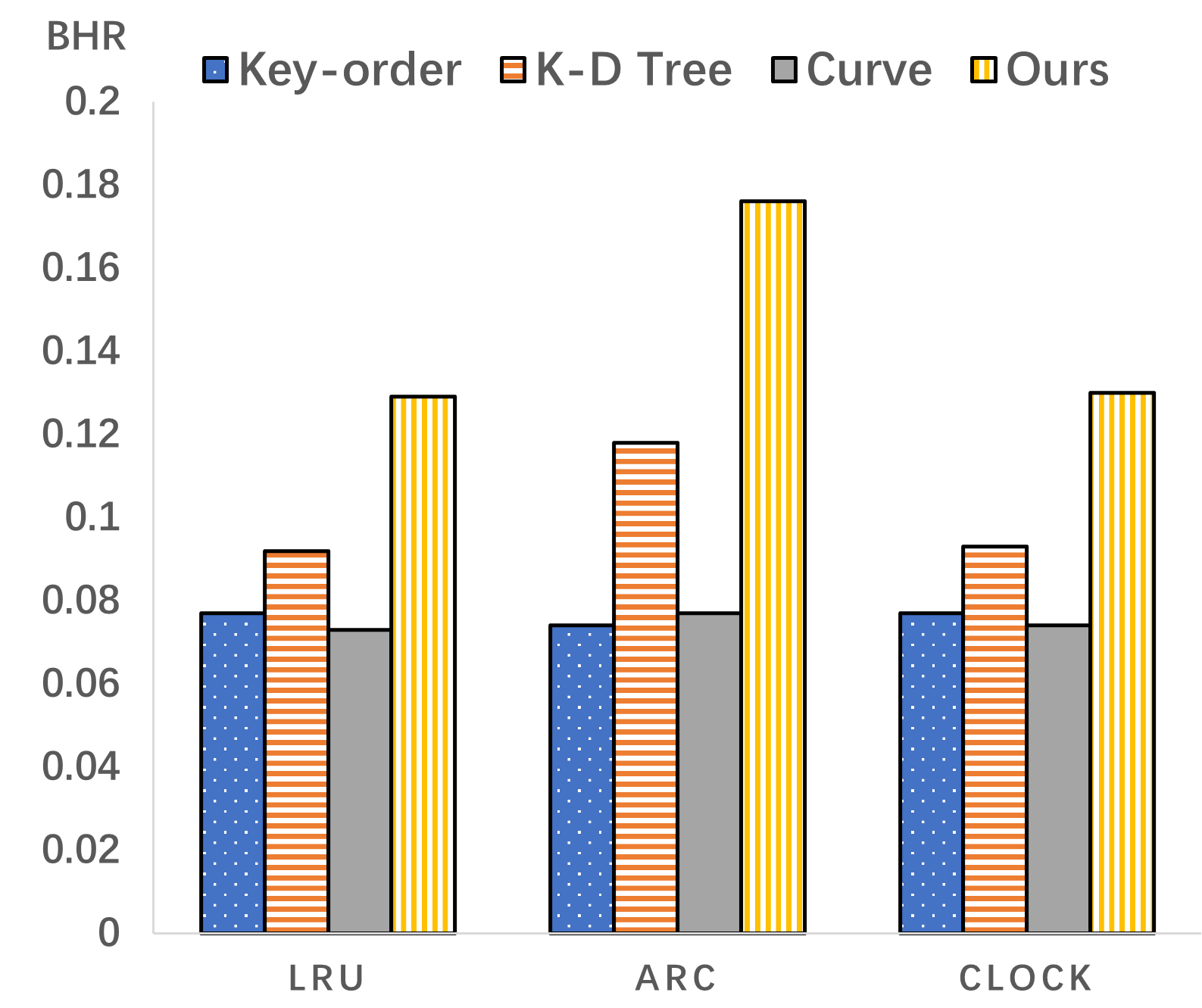 }}
        \vspace{-0.5em}
	\caption{Experimental Results of Using Different Cache Replace Strategy in Cloud-Edge-Device Cache Replacement Task }
         \label{p8}
         \vspace{-0.5em}
\end{figure}

\textbf{Effect of Different Cache Replace Strategy}: In this part, we focus on the impact of cache replacement strategies used by end devices on the cloud-edge-device cache replacement. Here, we fix the end cache size to 2\% of the dataset's scale, and the results are shown in \autoref{p8}. We observe that the Key-order and the Curve baseline are insensitive to the cache replacement strategy. In contrast, our method and the K-D Tree exhibit sensitivity to the cache replacement strategy. When using ARC for cache replacement, our method and the K-D Tree demonstrate optimal performance in both $THR$ and $BHR$ metrics. However, regardless of the cache replacement strategy employed, our method consistently outperforms others.

\section{Related Work} \label{sec:rel}

\parindent=0pt
\textbf{Hierarchical Storage}: Current research typically classifies data into two tiers based on hotness and coldness, focusing on mechanisms to differentiate between them. Hot data is frequently accessed, while cold data is accessed less often. Methods for classification fall into two groups: one relies on the characteristics of data structures, determining hotness or coldness based on the relative position of data within the data structure. Examples of this category include LRU~\cite{Chang2002AnAS}~\cite{ONeil1993TheLP} and ARC~\cite{MegiddoProceedingsOF}~\cite{Jiang2005DULOAE}~\cite{Cheng2015AMCAA}, commonly employed in cache management for operating systems. The other employs statistical methods to measure data hotness. By defining relevant metrics qualitatively or quantitatively, the heat of data is ascertained. ~\cite{Levandoski2013IdentifyingHA} adopts such approach, suggesting to use exponential smoothing algorithms on log access records to predict the likelihood of future data access.
In contrast, \textbf{Brame} classifies the data into three tiers to satisfy the diverse storage requirements of \textbf{CEDC}.

\parindent=0pt
\textbf{Data Migration and Scheduling}: Data migration can be categorized into real-time replacement and periodic migration scheduling. The former, known as cache replacement strategies, has been a long-standing focus, with classical methods such as LRU~\cite{Chang2002AnAS}, LFU~\cite{Karakostas2002ExploitationOD}~\cite{Jayarekha2010AnAD}, ARC~\cite{MegiddoProceedingsOF}, CLOCK~\cite{Jiang2005CLOCKProAE}~\cite{Lee2015MCLOCKMP}., have demonstrated significant success and widespread application over the past few decades. Recently, AI techniques like reinforcement learning (LeCaR~\cite{Vietri2018DrivingCR}) and GBDT-based methods (LRB~\cite{Song2020LearningRB}, MAT~\cite{Yang2023ALC}) have been explored for data eviction. Research on periodic migration is more limited , with relevant works include Gorilla~\cite{Pelkonen2015GorillaAF}, Xie's work~\cite{Xie2019}, and TS-Cabinet~\cite{Cui2023TSCabinetHS}. Gorilla retains data collected by the system over the last 26 hours as hot data in the cache and periodically removes outdated data to the disk. ~\cite{Xie2019} proposes quantitatively modeling the data temperature using Newton's cooling law and utilizes the high-low watermark method to eliminate cold data from hot storage media to cold media. TS-Cabinet, in the context of time-series databases, addresses the issue of hierarchical data placement. It suggests modeling data heat by combining Newton's cooling law and the Stefan-Boltzmann law and employs a method akin to ~\cite{Xie2019} to migrate data between the cold layer and the hot layer.
On the contrary, \textbf{Brame} uses $Blocks$ as the fundamental unit for data migration, provides a unified framework that supports both periodic migration scheduling between the cloud and edge and real-time cache replacement on the end.

%\vspace{-0.5em}
\section{Conclusions} \label{sec:con}
%\vspace{-0.5em}

This paper proposes \textbf{Brame}, a workload-aware three-tiered data storage framework for \textbf{CEDC}. By assigning specific storage roles to cloud, edge, and end devices, \textbf{Brame} optimizes cost reduction without sacrificing performance. \textbf{Brame} takes $Blocks$ as the basic unit for data management, introducing an offline block generation method to streamline data organization and query routing. \textbf{Brame} involves two key components: workload-aware data reorganization and query-driven partitioning. Extensive experiments, especially on downstream tasks, validate the feasibility and effectiveness of block-level data management. Results show the superior performance of \textbf{Brame}'s block generation component. Future work will integrate offline block generation with online tuning and extend the method to multi-table scenarios.

\bibliographystyle{ieeetr}
\bibliography{main}

\end{document}